\documentclass[aps,prd,twocolumn,superscriptaddress,groupedaddress,nofootinbib]{revtex4}
\usepackage{graphicx}  
\usepackage{dcolumn}   
\usepackage{bm}        
\usepackage{amssymb}   
\usepackage{amsmath}
\usepackage{appendix}
\usepackage[svgnames]{xcolor} 
\allowdisplaybreaks

\usepackage[colorlinks=True, linkcolor=SteelBlue, citecolor=SteelBlue,urlcolor=SteelBlue]{hyperref}
\newcommand{\mbf}[1]{\mathbf{#1}}
\usepackage{orcidlink}

\begin{document}

\title{Resonance locking: radian-level phase shifts due to nonlinear hydrodynamics of $g$-modes in merging neutron star binaries}
\author{K.J. Kwon\,\orcidlink{0000-0001-9802-362X}}
\affiliation{Department of Physics, University of California, Santa Barbara, CA 93106, USA}
\author{Hang Yu\,\orcidlink{0000-0002-6011-6190}}
\affiliation{eXtreme Gravity Institute, Department of Physics, Montana State University, Bozeman, MT 59717, USA}
\author{Tejaswi Venumadhav\,\orcidlink{0000-0002-1661-2138}}
\affiliation{Department of Physics, University of California, Santa Barbara, CA 93106, USA}
\affiliation{International Centre for Theoretical Sciences, Tata Institute of Fundamental Research, Bangalore 560089, India}
\date{\today}

\defcitealias{nlrl_prl}{KYV I}

\begin{abstract}
A neutron star~(NS) in a binary system deforms due to the companion's tidal gravitational field.
As the binary inspirals due to gravitational wave (GW) emission, the NS's deformation evolves; this evolution is typically modeled as the star's linear response to the companion's time-evolving tidal potential. 
In principle, the fluid elements' displacements can be excited and evolve nonlinearly since the equations of hydrodynamics and the tidal forcing have nonlinear terms. 
Recently, Kwon, Yu, and Venumadhav~(KYV~I) showed that nonlinear terms in the hydrodynamic equations of motion make the low-frequency response of NSs, characterized by gravity ($g$-) modes, behave in an anharmonic manner.
The anharmonicity is dominantly generated by the mutual coupling of the four lowest-order ($n=1$, $l=|m|=2$) $g$-modes, and allows them to stay locked in a resonant state that oscillates phase-coherently with the orbit throughout the inspiral. 
As a result, the $g$-modes grow to larger amplitudes than the linear response suggests, leading to an extra phase correction to the frequency-domain GW signal $|\Delta \Psi|\approx 3\,{\rm rad}$ at a GW frequency of $1.05\,{\rm kHz}$. 
This effect is part of the truly dynamical tide, in the sense that the amplitude depends not just on the binary's instantaneous frequency but the entire history of the inspiral.
In this paper, we explain the phenomenology of resonance locking in detail and analytically validate the numerical dephasing calculations in KYV~I. 
We also demonstrate that the effect is only significant for the lowest-order $g$-modes. 
\end{abstract}

\maketitle

\section{Introduction}
The first detection of gravitational waves~(GWs) from a binary neutron star~(BNS) merger, GW170817~\cite{gw170817}, enabled the first constraints on the nature of nuclear matter at supranuclear densities from a GW signal~\cite{gw170817_eos, gw170817_improve}. 
The event was accompanied by a gamma-ray burst~(GRB170817A)~\cite{grb170817a_ligo, grb170817a_fermi, grb170817a_integral} and kilonova~(AT2017gfo)~\cite{at2017gfo_ligo, at2017gfo_alexander, at2017gfo_arcavi,  at2017gfo_chornock, at2017gfo_coulter, at2017gfo_cowperthwaite, at2017gfo_drout, at2017gfo_evans, at2017gfo_hallinan, at2017gfo_nicholl, at2017gfo_smartt, at2017gfo_soares-santos, at2017gfo_tanvir}, which opened the door to multi-messenger astronomy with GW signals. Joint gravitational and electromagnetic~(EM) observations of neutron star~(NS) merger events provide measurements of stellar masses, radii, and ejecta properties, which can constrain the nuclear equation of state~(EOS)~\cite{gw170817_improve, gw170817_eos, radice2017a}. 

The impact of the NSs' internal structure on the GW signal arises from the stars' deformation within the companion's tidal field~\cite{lai94a, lai94b, kokkotas95, andersson98, mora04, flanagan08, damour09, read09, hinderer10, bini14, bernuzzi15,steinhoff16, hinderer16, ma21, steinhoff21, kuan22, kuan23, yu24}. 
Mathematically, the primary develops a quadrupole moment $Q_{ij}$ in response to a tidal field $\mathcal{E}_{ij}$. 
In the static limit, the quadrupole moment is linearly proportional to the instantaneous tidal field, and the ratio between the two quantities defines the tidal deformability $\lambda$, \emph{i.e.}, $Q_{ij}=-\lambda \, \mathcal{E}_{ij}$~(and its dimensionless version $\Lambda\equiv  \lambda M^{-5}c^{10}G^{-4}$ with $M$ being the mass of the primary)~\cite{flanagan08} which is an EOS-dependent quantity~\cite{hinderer08, hinderer08_erratum, binnington09, damour09, hinderer10}.
In reality, the NSs are orbiting each other and inspiraling due to the GW emission; 
when the time dependence is treated in an adiabatic manner, the nonzero $\lambda$ formally modifies the phasing of waveforms at a relatively high~($\geq 5$) post-Newtonian~(PN) order. 
Despite this, the tidal signature of the GW signals is potentially measurable due to the prefactor of the phase correction being numerically large. 

The data from GW170817 did not yield decisive evidence for a tidal phase shift; it led to an upper bound $\Lambda \leq 800$ for a $1.4\,M_{\odot}$ NS assuming a low-spin prior~\cite{gw170817, gw170817_improve}. 
The prospect of detecting more GWs from BNS systems~\cite{lvk_prospect_bns}, and the potential development of next-generation detectors~\cite{lisa, cosmic_explorer, einstein_telescope} that could be sensitive to the early and late stages of inspirals, motivates us to improve our theoretical understanding of the tidal dynamics of NSs.

In general, any displacement field in an NS can be decomposed as a sum of eigenmodes that describe its stellar oscillations; this basis simplifies the calculations of the tidal response of an NS, which is dominated by the fundamental oscillation~($f$-) modes that have the strongest coupling to the tidal potential \cite{lindbolm83, andersson98, ma20, kuan22, yu23, yu24}. 
Low-order $f$-modes characterize the large-scale distortion of NSs; their eigenfrequency approximately matches the inverse of the star's dynamical timescale. 
For nonspinning NS binaries considered in this paper~(see, \emph{e.g.}, Refs.~\cite{ho99, doneva13, ma20, yu24, yu25} for oscillations in rotating NSs), the tidal forcing frequency remains smaller than the $f$-modes' eigenfrequency, which makes them nearly adiabatic. 
The resulting dephasing of waveforms with respect to those from point-particle inspirals can largely be explained using $f$-modes and the linear theory.
Pressure~($p$-) modes also behave adiabatically as their eigenfrequencies are even higher than those of $f$-modes. Nonetheless, they alone do not significantly affect the inspiral due to their weak couplings to tides.

The tidal deformation also has corrections for finite tidal forcing frequency~\cite{steinhoff16, hinderer16, steinhoff21, gamba23} and dynamical components that depend on the history of tidal forcing \cite{lai94c, yu24, yu25}.\footnote{{We can distinguish adiabatic, equilibrium, and dynamical tides using the convention of Yu~\emph{et~al.}~\cite{yu24}; the adiabatic tide is the star's tidal response in the limit of zero tidal forcing frequency [Eqs.~(\ref{eqn:full_eom}) and (\ref{eq:drivingterms}), with ${\rm d^2}\chi_a/{\rm d}t=0$, and $\Phi=0$.]
The equilibrium tide admits a correction due to the finite tidal forcing frequency but depends only on the frequency's instantaneous value [\emph{e.g.}, Eqs.~(\ref{eqn:approx_sol0})--(\ref{eqn:approx_sol})].
Finally, the dynamical tide is the solution of the full initial-value problem, given by Eqs.~(\ref{eqn:full_eom}) and (\ref{eq:drivingterms}), which explicitly depends on the initial condition (and thus, the entire history of the inspiral).}} 
The dynamical effects arise from the fact that modes have finite eigenfrequencies. These effects can become significant at the later stage of the inspiral when the orbital frequency approaches the modes' eigenfrequencies.
For the $l=|m|=2$ $f$-modes, this results in a slight enhancement in the mode amplitude unless NSs are rotating, in which case $f$-modes can evolve through resonance. Gravity~($g$-) modes have eigenfrequencies that are low enough~\cite{reisenneger92} that the tidal forcing frequency can evolve through resonance even if the binaries are nonspinning. 
Under the assumption that the $g$-modes are linearly driven, this leads to resonant excitations that oscillate at the modes' eigenfrequencies. 
These excitations are phase-incoherent with respect to the orbit and have negligible impact on the GW signal past the resonance~\cite{lai94c}. 

Recently, Kwon,~Yu,~and~Venumadhav~(hereafter \citetalias{nlrl_prl})~\cite{nlrl_prl} demonstrated that the $g$-modes behave anharmonically~(\emph{i.e.}, their oscillation frequencies depend on their energies \cite{landau82, yu21, yu23}) once their nonlinear interactions are accounted for. 
If the $g$-modes are excited to sufficient energies for their nonlinearity to become important, the modes do not evolve through resonance and instead stay locked in a near-resonant state until the merger. 
This effect, dubbed resonance locking~(RL), is caused by the anharmonic frequency shift evolving in lock-step with the driving frequency. 
It is analogous to RLs due to spin evolution in inspiraling double white dwarfs~\cite{fuller12, burkart13, yu20} and structural evolution of exoplanet-hosting stars~\cite{ma21} (see also Burkart \emph{et~al.}~\cite{burkart14} for a general discussion on RL).

Whether or not RL can initiate in BNSs depends on the size of the effective four-mode coupling coefficient between the four resonant $g$-modes, $\kappa_{\rm eff}$.
Moreover, the imaginary component of the frequency shift due to the $f$- and $g$-modes~(parameterized using a coefficient $\gamma_{\rm eff}$) can extend the lock's duration, without which the lock would break due to the GW-induced damping.
Due to these two effects, the locked $g$-modes evolve phase-coherently with the orbit throughout and cause a significantly larger phase shift on GW signals than what is expected from the linearly driven $g$-modes. 
Using a $1.4$-$1.4\, M_\odot$ Newtonian NS binary constructed using the SLy4 EOS \cite{chabanat98}, \citetalias{nlrl_prl} found an extra phase shift for the frequency-domain GW signal of approximately $\mathcal{O}(3)\,{\rm rad}$ at $1.05\,{\rm kHz}$.

Calculating the exact phase shift would require detailed corrections for general relativistic effects, yet, the results presented by \citetalias{nlrl_prl} have significant implications. 
Numerical-relativity~(NR) simulations can probe tidal effects by solving the Einstein equations~\cite{hotokezaka13, hotokezaka15, dietrich17d, foucart19}.
NR simulations initialize NSs in quasi-equilibrium configurations and simulate the last $\mathcal{O}(10)$ orbits before the merger. 
RL is a dynamical process that happens early in the inspiral and depends on the history of the tidal forcing, and hence the initial conditions need to be revised for late-inspiral simulations to be able to capture the effect.

In this paper, we provide a detailed calculation of the RL effect that~\citetalias{nlrl_prl} described at a high level. 
The structure of this paper is as follows: in Sec.~\ref{sec:form}, we provide a brief overview of the formalism that allows us to describe the tidal deformation in terms of eigenmodes.
In Sec.~\ref{sec:res_lock}, we derive the effective coupling coefficients $\kappa_{\rm eff}$  and $\gamma_{\rm eff}$ that~\citetalias{nlrl_prl} use to explain RL. 
We show that RL is significant only for the lowest-order $g$-modes. 
In addition, we analytically estimate the impact of the locked $g$-modes of the GW phase on the frequency domain, which is in agreement with the numerical result presented by \citetalias{nlrl_prl}. 
In Sec.~\ref{sec:disc}, we explore whether the shear motion at the crust-core can damp the RL and find it is unlikely. We conclude in Sec.~\ref{sec:conc}.

\section{Formalism\label{sec:form}}
An NS in binary deforms due to the tidal gravitational field generated by the companion star. 
If we describe the deformation by a vector field $\bm{\chi}$ defined on the unperturbed background star, it satisfies an equation of motion that schematically looks like
\begin{align}
    \frac{\partial^2\bm{\chi}}{\partial t^2} + \mbf{C}[\bm{\chi}]= \sum_{i=1} \mbf{a}_{\rm ext}^{(i)}+\sum_{i=2}\mbf{a}^{(i)}_{\rm int},\label{eqn:eom_chi}
\end{align}
where $\mbf{C}$ is a linear operator describing the star's hydrodynamic response to perturbations, and $\mbf{a}_{\rm ext}^{(i)}$ and $\mbf{a}_{\rm int}^{(i)}$ are respectively the external and internal accelerations. 
The superscripts $(i)$ on the accelerations in Eq.~\eqref{eqn:eom_chi} indicate that the terms are of the $i$-th order in displacement $\bm{\chi}$ and the expansion parameter $(R/D)^{l+1}$ [see Eq.~(\ref{eqn:tidal_u})], where $R$ is the radius of the primary NS, $D$ is the orbital separation between the binary components, and $l$ is the angular quantum number of the tidal driving.
We review the derivation of the equation of motion, Eq.~\eqref{eqn:eom_chi}, in Appendix~\ref{app:nonlin}. 
Note that the term $\mbf{C}[\bm{\chi}]$ corresponds to the leading-order internal acceleration $\mbf{a}^{(1)}_{\rm int}$; the explicit functional form of the operator $\mbf{C}$ is given in \emph{e.g.}, Lynden-Bell~and~Ostriker~\cite{lyndenbell1967} and Schenk \emph{et~al.}~\cite{schenk02}. 

In this paper, we include terms up to $\mbf{a}^{(3)}_{\rm int}$ and $\mbf{a}^{(2)}_{\rm ext}$ on the right-hand side of Eq.~(\ref{eqn:eom_chi}).\footnote{As we state later, we only consider $l=2$ tidal potential, for which $\mbf{a}^{(n)}_{\rm ext}$ vanishes for $n\geq 3$.} 
We work in the basis of the \emph{oscillation eigenmodes} instead of the coordinate basis. The transformation between these bases involves expressing the displacement field $\bm{\chi}$ in terms of the eigenmodes as
\begin{align}
    \bm{\chi}(t,\mbf{x})=\sum_a \chi_a(t)\bm{\xi}_a(\mbf{x}),
\end{align}
where $\chi_a(t)$ is the amplitude of a mode $a$ labeled by quantum numbers $(n_a, l_a, m_a)$ representing radial, polar, and azimuthal orders. 
Each mode is associated with a spatial eigenfunction $\bm{\xi}_a$ and eigenfrequency $\omega_a$.
As the NSs are not rotating, the eigenmodes satisfy $\mbf{C}[\bm{\xi}_a]=\omega_a^2\bm{\xi}_a$. 
In addition, the eigenmodes obey orthogonality relations; we choose the following normalization:
\begin{align}
    \langle\bm{\xi}_a, \bm{\xi}_b\rangle=\int {\rm d}^3x\,\rho \, \bm{\xi}_a^\ast \cdot \bm{\xi}_b=\frac{E_0\delta_{ab}}{\omega_a^2},&&E_0=\frac{GM^2}{R},
\end{align}
where $M$ is the mass of the primary NS and $G$ is the gravitational constant. 
We review the procedure to compute the eigenmodes in Appendix~\ref{app:eigen}. 

In the eigenmode basis, Eq.~(\ref{eqn:eom_chi}) reduces to an ordinary differential equation 
\begin{align}
    &\frac{{\rm d}^2\chi_a}{{\rm d}t^2}+\omega_a^2 \chi_a = \frac{\omega_a^2}{E_0}\langle \bm{\xi}_a, \mbf{a}^{(2)}_{\rm int} + \mbf{a}^{(3)}_{\rm int} + \mbf{a}_{\rm ext}^{(1)}+\mbf{a}_{\rm ext}^{(2)}\rangle.\label{eqn:full_eom}
\end{align}
On this basis, the driving terms evaluate to
\begin{subequations}
\label{eq:drivingterms}
\begin{align}
    &\frac{\langle \bm{\xi}_a ,
    \mbf{a}_{\rm ext}^{(1)}\rangle}{E_0} =\frac{M'}{M}W_{lm}I_{a}\left(\frac{R}{D}\right)^{l+1}e^{-im\Phi}, \label{eqn:lin_drive}\\
    &\frac{\langle \bm{\xi}_a ,
    \mbf{a}_{\rm ext}^{(2)}\rangle}{E_0} =\frac{M'}{M}\sum_{blm}W_{lm}^\ast J_{ablm}^\ast \chi_b^\ast \left(\frac{R}{D}\right)^{l+1}e^{im\Phi},\label{eqn:nonlin_drive}\\
    &\frac{\langle \bm{\xi}_a ,
    \mbf{a}_{\rm int}^{(2)}\rangle}{E_0}=\sum_{bc}\kappa_{abc}\chi_b^\ast \chi_c^\ast,\label{eqn:a_chi2}\\
    &\frac{\langle \bm{\xi}_a,
    \mbf{a}_{\rm int}^{(3)}\rangle}{E_0}=\sum_{bcd}\kappa_{abcd}\chi_b^\ast \chi_c^\ast\chi_d^\ast,\label{eqn:a_chi3}
\end{align}
\end{subequations}
where $M'$ is the companion mass, $W_{lm}=4\pi(2l+1)^{-1} Y_{lm}(\pi/2,0)$, $m$ is the azimuthal order of spherical harmonics, and $\Phi=\int {\rm d}t\,\Omega$ is the orbital phase defined with respect to the orbital frequency $\Omega$. 
The expressions in Eqs.~\eqref{eq:drivingterms} involve extra coefficients:  the linear and quadratic tidal coupling coefficients $I_a$ and $J_{ablm}$ are real-valued and can be calculated as~\cite{w12}
\begin{align}
    &I_{a}=\frac{1}{MR^l}\int {\rm d}^3x \rho \,\bm{\xi}_a^\ast \cdot \nabla (r^l Y_{lm}),\label{eqn:lin_coup}\\
    &J_{ablm}=\frac{1}{MR^l}\int{\rm d}^3x\rho\, \bm{\xi}_a \cdot (\bm{\xi}_b \cdot \nabla)\nabla (r^l Y_{lm}).\label{eqn:quad_coup}
\end{align}
The three- and four-mode coupling coefficients, $\kappa_{abc}$ and $\kappa_{abcd}$, are also real-valued and their expressions can be found in~\cite{vhoolst94, wu01, schenk02, w12, vzh, w16}. 
Eqs.~(\ref{eqn:lin_coup})~and~(\ref{eqn:eigenmode}) imply that only the eigenmodes with the same angular quantum numbers as the tidal potential are linearly forced. 
In contrast, $J_{ablm}$ allows modes to have different angular quantum numbers than the tidal potential~\cite{w12}. 
The modes in the three- and four-mode interactions $\kappa_{abc}$ and $\kappa_{abcd}$ must respectively satisfy $m_a + m_b + m_c = 0$ and $m_a + m_b + m_c + m_d=0$.

When we numerically solve for the eigenmodes (following the procedure in Appendix~\ref{app:eigen}) and the coupling coefficients, we have to choose an appropriate equation of state and solve for the star's background structure. 
We construct a Newtonian NS of mass $M=1.4\,M_\odot$ and radius $R\approx 13\,{\rm km}$ using the SLy4 equation of state~\cite{chabanat98}.
We assume that the NS is cold and consists of a normal fluid of neutrons, protons, and electrons.
We solve for $\beta$-equilibrium at each radial coordinate, and the resulting compositional gradient provides buoyancy that supports $g$-modes. 
We set the Brunt-V\"ais\"al\"a frequency to zero when the baryon number density of the NS is smaller than $0.08 \times 10^{39}\ {\rm fm}^{-3}$.
This treatment is consistent with studies that estimated the density at the inner crust boundary (\emph{e.g.}, \cite{douchin00}). 
We adopt the Cowling approximation for the $g$-modes~\cite{cowling41}.
This does not affect the accuracy of our analysis, since the $g$-modes weakly perturb the gravitational potential of the NS~\cite{finn88}.
When driving the eigenmodes, we only consider the $l=2$ harmonics of the tidal potential. 

In our computation, we include $g$-modes with the radial order $n=1$ and angular order $l=2$, as well as the $l=0,2,4$ $f$- and $p$-modes.
We consider $p$-modes with radial orders up to $n=60$, which covers the highest-order modes with eigenfrequencies lower than the acoustic cutoff limit of our NS.
\begin{figure}
    \centering
    \includegraphics[width=0.48\textwidth]{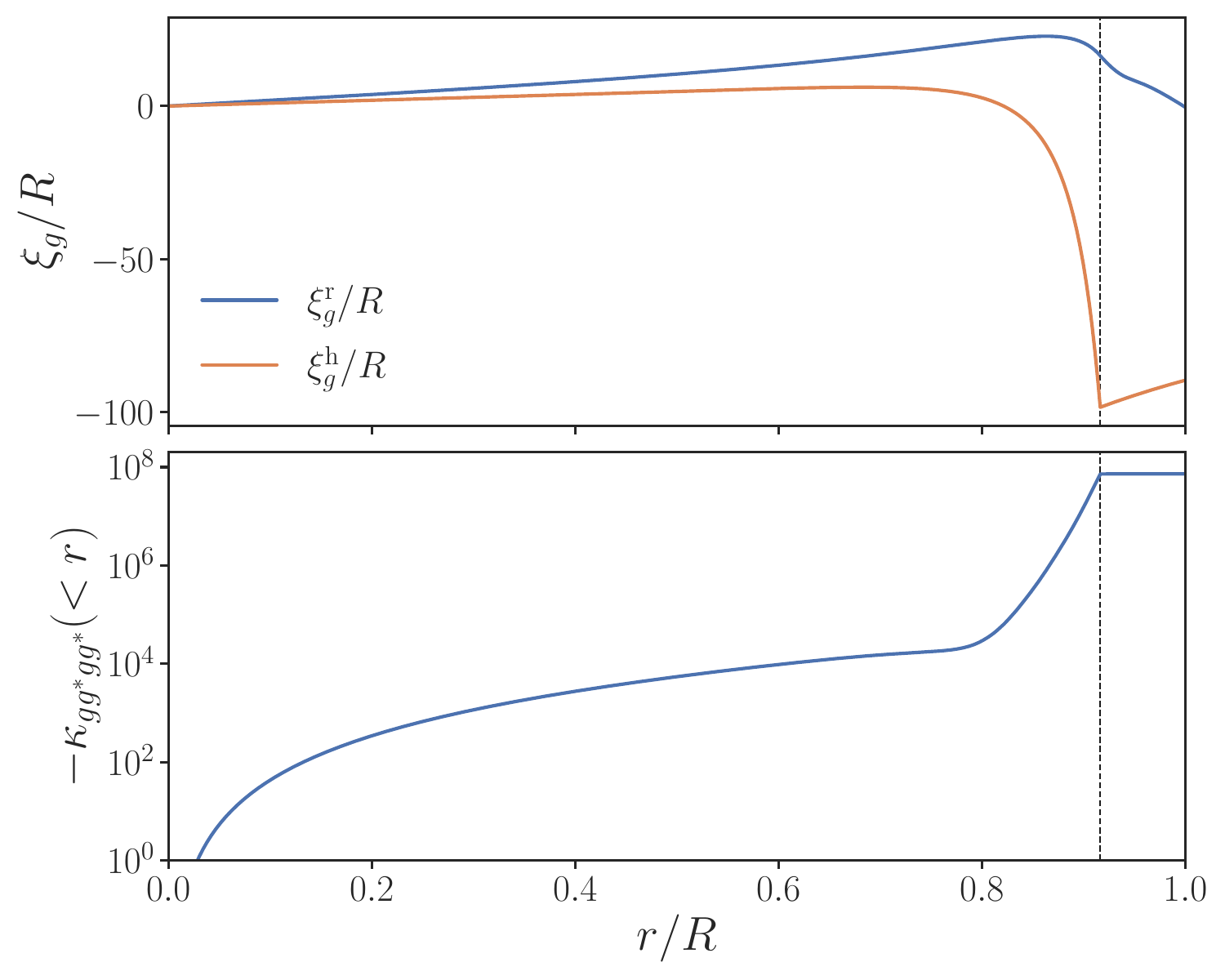}
    \caption{\textit{Top}---The eigenfunction of $g$-modes of radial order $n=1$ and angular order $l=2$. We show the radial (blue) and transverse (orange) components of the eigenfunction.
    The $g$-mode becomes evanescent within the crust, of which the inner boundary is marked with the vertical dashed line. \textit{Bottom}---The negative of the four-mode coupling profile calculated for two pairs of $n=1$, $l=2$, $m=\pm 2$ $g$-modes.
    The dominant accumulation of the coupling profile occurs between $0.8\lesssim r/R\lesssim 0.9$.
    \label{fig:mode}}
\end{figure}

In the top panel of Fig.~\ref{fig:mode}, we show the eigenfunction of the $n=1$, $l=2$ $g$-mode.
Using the black dashed line, we also show the inner boundary of the crust beyond which the $g$-mode becomes evanescent.
In the bottom panel, we show the four-mode coupling profile between two pairs of resonant $n=1$, $l=2$ $g$-mode. 
We see that the dominant contribution to the four-mode coupling occurs between $0.8\lesssim r/R\lesssim 0.9$.  

When we numerically model the inspiral, we dynamically evolve only the $l=2$ $f$- and $g$-modes, and assume the other modes are adiabatically sourced, \emph{i.e.}, $\ddot{\chi}=0$ in Eq.~\eqref{eqn:full_eom} for the other modes. 
We provide more details on the numerical integration in Sec.~\ref{subsec:ham}.

\section{Resonance Locking\label{sec:res_lock}}
\citetalias{nlrl_prl} provided a high-level description of RL of $g$-modes in a coalescing $1.4$-$1.4\, M_{\odot}$ NS binary:
The anharmonic $g$-mode shifts its frequency up as its energy increases. 
In turn, this frequency shift modifies the energy of the mode by tuning its resonance with the tidal forcing.
Thus, the anharmonicity creates a ``feedback loop'' that locks the evolution of mode frequency and energy to the orbit.
In this section, we provide more details about this phenomenology. 

In the interest of quickly making a connection with the calculations in \citetalias{nlrl_prl}, we will begin with a direct derivation of the effective coupling and damping coefficients $\kappa_{\rm eff}$ and $\gamma_{\rm eff}$ that are used in \citetalias{nlrl_prl} (and discard other terms) and then walk through the phenomenology of RL in Sec.~\ref{subsec:simp}. 
In Sec.~\ref{subsec:ham}, we will step back and write out the Hamiltonian of the binary and derive the full equations of motion that can be numerically evolved through the inspiral to compare to the intuitive `effective couplings' approach of Sec.~\ref{subsec:simp}. 
Finally, we present the numerically modeled inspiral and analytically validate the dephasing of the frequency-domain GW signal.

\subsection{Simple Description of RL Using $\kappa_{\rm eff}$ and $\gamma_{\rm eff}$\label{subsec:simp}}

\subsubsection{Equation of motion}
In this section, we derive a simplified equation of motion for anharmonic $g$-modes that captures the phenomenology of RL.
For a concise demonstration, we will continue our discussion from Eqs.~\eqref{eqn:full_eom} and \eqref{eq:drivingterms}.
We provide a more complete description of how we construct equations of motion using Hamiltonians in Sec.~\ref{subsec:ham}.

The full equations of motion, Eqs.~\eqref{eqn:full_eom} and \eqref{eq:drivingterms}, contain terms involving amplitudes of all the modes in the star. 
Our approach in this section is to identify those terms that significantly contribute to the RL of the $g$-modes. 
To form a feedback loop that can lead to RL, the fractional anharmonic frequency shift needs to depend on the $g$-mode amplitude. 
Therefore, it requires the $g$-mode to enter the coupling at least three times (one so that the coupling is about the $g$-mode, one so that an anharmonic frequency shift is created, and another one to form the frequency-energy feedback loop for locking). 
Meanwhile, conservation of angular momentum (selection rule of modes) forbids the direct coupling of three $|m|=2$ $g$-modes (the dominant nonlinear effect in Yu \emph{et~al.}~\cite{yu23} does not enter here). 
Hence, the coupling of interest has to involve four modes: three resonant $g$-modes and one mode $x$. 
We further restrict our attention on $x$ to those that have the most significant contributions either due to a temporal resonance or a strong spatial coupling. 
The former means the mode $x$ is another resonantly excited g-mode (we use $L_g$ to describe the Lorentzian-like resonant amplification of the $g$-mode), while in the latter, $x$ is an $f$ mode [we use $I_f$ for its spatial overlap; see Eq.~\eqref{eqn:lin_coup}]. 
$L_g$ and $I_f$ are therefore large parameters to assist order counting.

Let us consider the $g$-mode with the radial order $n=1$ and the quantum number $l=m=2$, and denote its amplitude with $\chi_g$.
We can read off its equation of motion from Eqs.~(\ref{eqn:full_eom}) and \eqref{eq:drivingterms}:
\begin{align}
    \ddot{\chi}_g + \omega_g^2 \chi_g &= \omega_g^2\Bigg(U_g + \sum_a U_{ga}^\ast \chi_a^\ast+ \sum_{ab}\kappa_{gab}\chi_a^\ast\chi_b^\ast \nonumber \\
    &\phantom{=\omega_g^2\Bigg(} + \sum_{abc}\kappa_{gabc}\chi_a^\ast\chi_b^\ast\chi_c^\ast \Bigg),\label{eqn:g_eom}
\end{align}
Here, we have defined the coefficients
\begin{align}
  U_g & = \frac{M'}{M} I_{g} W_{22} \left( \frac{R}{D} \right)^3 e^{-2i\Phi} \ {\rm and} \label{eq:Ug} \\
  U_{ga}^\ast & = \frac{M'}{M} \sum_{m} J_{ga2m} W_{2m} \left( \frac{R}{D} \right)^3e ^{im\Phi}
\end{align}
which are simply the linear and nonlinear tidal driving, each given by Eqs.~(\ref{eqn:lin_drive}) and (\ref{eqn:nonlin_drive}), evaluated for the $g$-mode of interest; these terms depend on the mode eigenfunction's coupling coefficients [$I_g$ and $J_{ga2m}$ defined in Eqs.~(\ref{eqn:lin_coup}) and (\ref{eqn:quad_coup})] and on the tidal field strength set by the orbital separation and the orbital phase. 

In the right-hand side of Eq.~(\ref{eqn:g_eom}), any term that contains $\chi_g$ causes a nonlinear frequency shift of the $g$-mode. 
In particular, the terms $\propto |\chi_g|^2\chi_g$ induce frequency shifts $\propto |\chi_g|^2 \sim L_g^2$, which contains two factors of the resonant $g$-mode's amplitude.
As a result, the $g$-mode becomes anharmonic, since $|\chi_g|^2$ is proportional to its linear energy. 

At first glance, one might assume that extracting the relevant portion of the final expression in Eq.~\eqref{eqn:g_eom} suffices to estimate the size of the effect. 
However, this represents only a partial account of the anharmonic shift, since the $g$-mode can drive other daughter modes whose amplitudes, in turn, amplitudes contain `hidden' factors of $L_g$. 
In other words, the anharmonic shift can occur through two channels of four-mode interaction: (i) self-coupling of $g$-modes via direct four-mode interaction, and (ii) effective four-mode couplings formed by chaining two three-mode couplings mediated by high-frequency daughter modes. 
We will refer to the frequency shifts resulting from both channels as $gggg$ as they involve interaction between four $g$-modes.
These two channels dominate because they have the largest spatial coupling coefficients.

In the former channel, the direct four-mode interaction that directly couples the $g$-modes enters Eq.~(\ref{eqn:g_eom}) as 
\begin{align}
    \ddot{\chi}_g + \omega_g^2 \chi_g = 3\omega_g^2 \kappa_{gg^\ast g g^\ast}|\chi_g|^2\chi_g + \cdots,
\end{align}
or equivalently,
\begin{align}
    \ddot{\chi}_g + \omega_g^2 (1-3 \kappa_{gg^\ast g g^\ast}|\chi_g|^2)\chi_g =\cdots.
\end{align}
Thus, we identify the frequency shift $\omega_g^2\rightarrow \omega_g^2 + \Delta_r \omega_g^2$
\begin{align}
    \Bigg(\frac{\Delta_r \omega_g^2}{\omega_g^2}\Bigg)_{\rm (i)}=-3\kappa_{gg^\ast gg^\ast}|\chi_g|^2,\label{eqn:anharm1}
\end{align}
where the subscript $r$ of $\Delta_r$ means that the shift $\propto |\chi_g|^2$ is real. 
The $g$-mode is now anharmonic, as its frequency depends on $|\chi_g|^2$.

The second channel that makes the $g$-mode anharmonic is mediated by three-mode coupling terms between two low-order $g$-modes and high-frequency modes, which we refer to as $\beta$. 
These modes $\beta$ consist of $f$-modes with $l=0$ and 4 and $p$-modes with $l=0,2,$ and 4.
The second channel enters Eq.~(\ref{eqn:g_eom}) as
\begin{align}
&\ddot{\chi}_g + \omega_g^2\chi_g = \omega_g^2\Bigg(2\sum_\beta^{m_\beta=0}\kappa_{gg^\ast \beta}\chi_g \chi_\beta^\ast \nonumber\\
&\phantom{\ddot{\chi}_g + \omega_g^2\chi_g=\omega_g^2\Big(}+2\sum_\beta^{m_\beta={-4}}\kappa_{gg\beta}\chi_g^\ast \chi_\beta^\ast +\cdots\Bigg).\label{eqn:eff_four_demo}
\end{align} 
Given the selection rules for $\beta$, the corresponding eigenfunctions $\bm{\xi}_\beta$ couple weakly with the linear tidal drive, which lets us ignore Eq.~(\ref{eqn:lin_drive}) in the equations of motion of $\chi_\beta$.
Then, $\chi_\beta$ is driven by coupling across modes and the nonlinear tidal drive given by Eqs.~(\ref{eqn:nonlin_drive})--(\ref{eqn:a_chi3}).
Moreover, $\chi_\beta$'s natural frequencies are high enough that the adiabatic approximation applies for their amplitudes.
Considering only the nonlinear terms relevant to the frequency shift $\propto |\chi_g|^2 \sim L_g^2$, $\chi_\beta$ evaluates to
\begin{align}
    &\chi_\beta = 2\kappa_{gg^\ast \beta}|\chi_g|^2 +\cdots,&&(m_\beta=0),\label{eqn:chi_beta_eff_four_m0}\\
    &\chi_\beta = \kappa_{gg\beta}(\chi_g^\ast)^2 + \cdots,&&(m_\beta=-4).\label{eqn:chi_beta_eff_four_m-4}
\end{align}
Substituting these expressions for $\chi_\beta$ back into Eq.~(\ref{eqn:eff_four_demo}), we obtain
\begin{align}
    &\ddot{\chi}_g+\omega_g^2\chi_g = \omega_g^2\Bigg(4\sum_\beta^{m_\beta=0}\kappa_{gg^\ast\beta}^2|\chi_g|^2\chi_g\nonumber\\
    &\phantom{\ddot{\chi}_g+\omega_g^2\chi_g = \omega_g^2\Bigg(}+2\sum_\beta^{m_\beta=-4}\kappa_{gg\beta}^2|\chi_g|^2\chi_g+\cdots\Bigg).\label{eqn:exp_fshift}
\end{align}
The two terms on the right-hand side illustrate that the two three-mode couplings have been chained to form an effective four-mode coupling.
Moving these terms to the left-hand side, we have
\begin{align}
    &\ddot{\chi}_g + \omega_g^2\Bigg[1-\Bigg(4\sum_{\beta}^{m_\beta=0}\kappa_{gg^\ast\beta}^2 
    + 2\sum_{\beta}^{m_\beta=-4}\kappa_{gg\beta}^2\Bigg)|\chi_g|^2\Bigg]\chi_g\nonumber \\
    &=\cdots.
\end{align} 
This looks like an anharmonic shift due to an effective four-mode interaction 
\begin{align}
    &\Bigg(\frac{\Delta_r \omega_g^2}{\omega_g^2}\Bigg)_{\rm (ii)}=-\Bigg(
    4\sum_{\beta}\kappa_{gg^\ast\beta}^2+
    2\sum_{\beta}\kappa_{gg\beta}^2\Bigg)|\chi_g|^2,\label{eqn:anharm2}
\end{align}
Equations~(\ref{eqn:anharm1}) and (\ref{eqn:anharm2}) together fully describe the anharmonic frequency shift at the lowest order in $g$-mode energy $(\propto |\chi_g|^2 \sim L_g^2)$, which enters the equation of motion as
\begin{align}
    \ddot{\chi}_g + \omega_g^2\left[1+\Bigg(\frac{\Delta_r\omega_g^2}{\omega_g^2}\Bigg)_{\rm (i)}+\Bigg(\frac{\Delta_r\omega_g^2}{\omega_g^2}\Bigg)_{\rm (ii)}\right]\chi_g=\cdots.\label{eqn:kap_deriv}
\end{align}
These anharmonic terms play a crucial role in initiating RL~(\citetalias{nlrl_prl}). 
This motivates us to define an effective coupling coefficient that governs the $g$-mode's anharmonicity
\begin{align}
    \kappa_{\rm eff}&\equiv -\frac{1}{|\chi_g|^2}\left[\Bigg(\frac{\Delta_r \omega_g^2}{\omega_g^2}\Bigg)_{\rm (i)}+\Bigg(\frac{\Delta_r\omega_g^2}{\omega_g^2}\Bigg)_{\rm (ii)}\right]\nonumber\\
    &=3\kappa_{gg^\ast gg^\ast}+4\sum_\beta^{m_\beta=0} \kappa_{gg^\ast \beta}^2 + 2\sum_\beta^{m_\beta=-4} \kappa_{gg\beta}^2,\label{eqn:k_eff_0}
\end{align}
and to rewrite Eq.~(\ref{eqn:kap_deriv}) as
\begin{align}
    \ddot{\chi}_g + \omega_g^2(1-\kappa_{\rm eff}|\chi_g|^2)\chi_g=\cdots.
\end{align}
When the $g$-mode evolves through the linear resonance point ($2\pi f_{\rm gw}=\omega_g$), it is near-resonantly driven by the tidal forcing. 
Near this point, the real part of the frequency shift is dominated by Eqs.~(\ref{eqn:anharm1}) and (\ref{eqn:anharm2}), \emph{i.e.}, ${\rm Re}[\Delta \omega_g^2]\sim -\omega_g^2\kappa_{\rm eff}|\chi_g|^2$. 
There is typically a large cancellation between the `bare' four-mode term $3 \, \kappa_{gg^\ast gg^\ast}$ and the adiabatically sourced terms involving $\beta$ in Eq.~\eqref{eqn:k_eff_0}, and hence it is important to include both channels.

Let us now consider shifts of the form $gggf$, which contain three low-order $g$-modes (with $|m|=2$) and one $f$-mode that has the largest spatial overlap with the tidal potential. 
Angular selection rules enforce that the $f$-mode has $l=|m|=2$. 
We will use $\chi_f\sim I_f(R/D)^3$ to denote the amplitude of the $l=m=2$ $f$-mode. 
This shift involves the four-mode interaction between three $g$-modes and one $f$-mode, and hence we will refer to it as $gggf$. 

The coupling coefficients relevant to the $gggf$ shift are much smaller than those that cause the $gggg$ shift. 
Hence, the frequency shift is dominated by the $gggg$ interaction as mentioned previously. 
However, the $gggf$ interaction introduces a small imaginary component to the frequency shift (which the $gggg$ terms do not contribute to).
This new imaginary frequency is the key result of the energy exchange between $f$- and $g$-modes, which modifies the dynamics of RL as we explain below.

Again, the formal four-mode coupling enters Eq.~(\ref{eqn:g_eom}) as
\begin{align}
\ddot{\chi}_g + \omega_g^2 \chi_g &= \omega_g^2(6\kappa_{ggg^\ast f^\ast}\chi_g^\ast \chi_f +3\kappa_{gg^\ast g^\ast f}\chi_g\chi_f^\ast)\chi_g + \cdots\nonumber\\
&=3i\omega_g^2 \kappa_{ggg^\ast f^\ast}{\rm Im}[\chi_g^\ast \chi_f]\chi_g+\cdots,\label{eqn:fggg_chan1}
\end{align}
where in the second equality we have neglected the real component of the frequency shift and used $\kappa_{ggg^\ast f^\ast}=\kappa_{gg^\ast g^\ast f}$ is real.
Therefore, we find the imaginary shift $\omega_g^2\rightarrow \omega_g^2 + \Delta_i\omega_g^2$
\begin{align}
    \left(\frac{\Delta_i \omega_g^2}{\omega_g^2}\right)_{\rm (i)}=-3i\kappa_{ggg^\ast f^\ast}{\rm Im}[\chi_g^\ast \chi_f].\label{eqn:im_shift1}
\end{align}

Just as in the case of the $gggg$ coupling, there is a second channel that contributes to the $gggf$ coupling in the same order, involving high-frequency modes $\beta$.
The driving equation for the $g$-mode due to this second channel looks like
\begin{align}
    \ddot{\chi}_g + \omega_g^2 \chi_g = \omega_g^2\Bigg[2\sum_{\beta}^{m_\beta=0}(\kappa_{gg^\ast \beta}\chi_g+\kappa_{gf^\ast \beta}\chi_f) \chi_\beta^\ast\nonumber \\
    +2\sum_{\beta}^{m_\beta=-4}(\kappa_{gf \beta}\chi_f^\ast +\kappa_{gg\beta}\chi_g^\ast) \chi_\beta^\ast+\sum_\beta U_{g\beta}^\ast \chi_\beta^\ast +\cdots\Bigg],\label{eqn:gamma_eff_0}
\end{align}
where we have included for the coupling across $\beta$, $f$\mbox{-,} and $g$-modes. 
We additionally include the nonlinear tidal driving terms $U_{g\beta}\propto (R/D)^3$, because they have the same scaling as the adiabatic $f$-mode amplitude $\chi_f\propto  (R/D)^3 \sim I_f$.
Although $m_\beta=-4$, $-2$, and $0$ are allowed,  $|m_\beta|=2$ does not contribute to the imaginary shift at the considered order. 
We solve for the amplitudes $\chi_\beta$ of the high-frequency modes using the adiabatic approximation as earlier, including the couplings with the $f$-modes and $U_{g\beta}$ on top of Eqs.~(\ref{eqn:chi_beta_eff_four_m0}) and (\ref{eqn:chi_beta_eff_four_m-4})
\begin{align}
    &\chi_\beta = 2\kappa_{gg^\ast \beta}|\chi_g|^2+2\kappa_{gf^\ast\beta}\chi_g^\ast \chi_f+2\kappa_{g^\ast f\beta}\chi_g\chi_f^\ast \nonumber \\
    &\phantom{\chi_\beta =}+U_{\beta g}^\ast \chi_g^\ast + U_{\beta g^\ast}^\ast \chi_g + \cdots,\ \ \ \ \ \ \ \ \ \text{$(m_\beta=0)$}, \label{eqn:adi_gamma_0}\\
    &\chi_\beta = \kappa_{gg \beta}(\chi_g^\ast)^2+2\kappa_{gf\beta}\chi_g^\ast \chi_f^\ast +U_{\beta g}^\ast \chi_g^\ast + \cdots,\nonumber  \\
    &\phantom{\phantom{\chi_\beta =}+U_{\beta g}^\ast \chi_g^\ast + U_{\beta g^\ast}^\ast \chi_g + \cdots,} \ \ \ \ \ \ \ \ \ \, \text{$(m_\beta=-4)$}.\label{eqn:adi_gamma_1}
\end{align}
The nonlinear tidal terms $U_{g\beta}$ in Eqs.~(\ref{eqn:adi_gamma_0}) and (\ref{eqn:adi_gamma_1}) cause shifts $\propto \chi_f^\ast \chi_g$ and $\propto \chi_f \chi_g^\ast$ (\emph{i.e.}, $\sim L_g I_f$) when we replace $\chi_\beta^\ast$ in the first and second summations in Eq.~(\ref{eqn:gamma_eff_0}).\footnote{A careful reader might wonder why we omitted $U_{g\beta}$ in the analogous equations~(\ref{eqn:chi_beta_eff_four_m0}) and (\ref{eqn:chi_beta_eff_four_m-4}). 
In detail, these terms are there, but the coefficients $U_{g\beta}$ do not contain factors of $L_g$ and cannot cause frequency shifts $\propto |\chi_g|^2 \sim L_g^2$, so we omitted them for brevity.}
Substituting the above expressions for $\chi_\beta$ back into Eq.~(\ref{eqn:gamma_eff_0}) and discarding terms that do not contribute to the imaginary shift at the considered order, we obtain
\begin{align}
    &\ddot{\chi}_g + \omega_g^2\chi_g =\omega_g^2\Bigg[\sum_{\beta}^{m_\beta=0}2\kappa_{gg^\ast \beta}(2\kappa_{g^\ast f\beta}\chi_g^\ast \chi_f+  U_{g\beta}^\ast\chi_g^\ast)\nonumber \\
    &+\sum_{\beta}^{m_\beta=-4}\kappa_{gg\beta}(2\kappa_{gf\beta}\chi_f\chi_g^\ast+U_{\beta g}\chi_g^\ast)\Bigg]\chi_g + \cdots.
\end{align}
Thus, we arrive at the expression for the imaginary shift due to the second channel
\begin{align}
    \left(\frac{\Delta_i \omega_g^2}{\omega_g^2}\right)_{\rm (ii)}=-i \, {\rm Im}\Bigg[\sum_{\beta}^{m_\beta=0}2\kappa_{gg^\ast \beta}\left(2\kappa_{gf^\ast \beta}\chi_f+U_{g\beta}^\ast \right)\chi_g^\ast\nonumber \\
    +\sum_{\beta}^{m_\beta=-4}\kappa_{gg \beta}\left( 2\kappa_{gf\beta}\chi_f + U_{\beta g}\right)\chi_g^\ast\Bigg].\label{eqn:im_shift2}
\end{align}
Equations~(\ref{eqn:im_shift1}) and (\ref{eqn:im_shift2}) shift the imaginary component of the frequency, behaving as an effective (anti-) damping of the $g$-modes.
Therefore, we define an effective (anti-) damping coefficient $\gamma_{\rm eff}$ such that
\begin{align}
    \gamma_{\rm eff}{\rm Im}[\chi_g\chi_f^\ast]&\equiv\left(\frac{\Delta_i \omega_g^2}{\omega_g^2}\right)_{\rm (i)}+\left(\frac{\Delta_i \omega_g^2}{\omega_g^2}\right)_{\rm(ii)}.
\end{align}
Treating $U_{g\beta}/\chi_f$ as a real-valued constant 
\begin{align}
    \frac{U_{g\beta}}{\chi_f}=\frac{J_{g\beta 22}W_{22}(R/D)^3}{I_fW_{22}(R/D)^3}=\frac{J_{g\beta22}}{I_f}
\end{align}
and using the fact that coupling coefficients are real-valued and symmetric upon conjugation of modes, the expression for $\gamma_{\rm eff}$ reads
\begin{align}
&\gamma_{\rm eff}=3\kappa_{gg g^\ast f^\ast}+ \sum_\beta^{m_\beta=0}2\kappa_{gg^\ast \beta} \left (2\kappa_{\beta g^\ast f} + \frac{U_{\beta g^\ast }^\ast}{\chi_f^\ast}\right)\nonumber \\
    &\phantom{\gamma_{\rm eff}=}\ +\sum^{m_\beta=-4}_\beta \kappa_{gg\beta}\left(2\kappa_{gf\beta}+\frac{U_{\beta g}^\ast}{\chi_f^\ast}\right).\label{eqn:gamma_eff}
\end{align}
The upshot is that the phenomenology of RL can be faithfully captured using only the coefficients $\kappa_{\rm eff}$ and $\gamma_{\rm eff}$ [in Eqs.~\eqref{eqn:k_eff_0} and \eqref{eqn:gamma_eff}], with a simplified equation of motion:
\begin{align}
    &\ddot{\chi}_g + \omega_{\rm eff}^2\chi_g =\omega_g^2U_g,\label{eqn:toy_eom}\\
    &\omega_{\rm eff}^2 = \omega_g^2(1-\kappa_{\rm eff}|\chi_g|^2 + i\gamma_{\rm eff}{\rm Im}[\chi_f^\ast \chi_g]),\label{eqn:toy_eff}
\end{align}
where $\omega_{\rm eff}$ is the effective frequency of the system. In Table~\ref{tab:eff_vals}, we summarize the values of $\kappa_{\rm eff}$, $\gamma_{\rm eff}$, and $\omega_g$ that enter Eq.~(\ref{eqn:toy_eom}).
We additionally include critical values for these coefficients, $\kappa_{\rm crit}$ and $\gamma_{\rm crit}$, relevant to the dynamics of RL, which we will derive in the following two subsections.
\begingroup
\renewcommand{\arraystretch}{1.2}
\begin{table}
\caption{Effective coupling coefficient $\kappa_{\rm eff}$, effective \mbox{(anti-)} damping coefficient $\gamma_{\rm eff}$, and eigenfrequency $\omega_g$ of the $g$-modes with radial order $n=1$ and angular order $l=2$. 
We additionally show the critical coupling coefficient and \mbox{(anti-)} damping coefficient, $\kappa_{\rm crit}$ and $\gamma_{\rm crit}$, which determine the initiation and the continuation of the lock.
Our $l=2$ $f$-modes and $n=1$, $l=2$ $g$-modes satisfy $I_f I_g < 0$.
Also, as we explain in Sec.~\ref{sec:res_lock}, whether a mode can enter and stay in a locked phase is not sensitive to the exact values of $\kappa_{\rm eff}$ and $\gamma_{\rm eff}$. 
Instead, there are ranges of values that $\kappa_{\rm eff}$ and $\gamma_{\rm eff}$ in which RL can occur. \label{tab:eff_vals} 
}
\begin{ruledtabular}
\begin{tabular}{ccccc}
$\kappa_{\rm eff}$&
$\kappa_{\rm crit}$&
$\gamma_{\rm eff}$&
$\gamma_{\rm crit}$&
$\omega_g/2\pi$\\
\colrule
$-2\times 10^7$ & $-1.5\times 10^6$ & $-1.0\times 10^5$ & $-2.0\times 10^4$ & $94.6\,{\rm Hz}$ \\
\end{tabular}
\end{ruledtabular}
\end{table}
\endgroup

\subsubsection{Initiation of the lock}
Based on Eqs.~(\ref{eqn:toy_eom}) and (\ref{eqn:toy_eff}), we can analytically estimate the evolution of the mode amplitude $\chi_g$. 
To this purpose, we factor out the rapidly accumulating orbital phase from $\chi_g$ and $U_g$ by defining 
\begin{align}
    X_g & = \chi_g e^{2i\Phi} \ {\rm and} \\
    V_g & = U_g e^{2i\Phi}.\label{eqn:vg}
\end{align}
Comparing $V_g\sim (R/D)^3$ with Eq.~\eqref{eq:Ug} tells us $V_g$ evolves on the radiation reaction timescale~$\sim (\dot{\Omega}/\Omega)^{-1}$ instead of the orbital timescale~$\sim \Omega^{-1}$. 
With these definitions, Eq.~(\ref{eqn:toy_eom}) reads
\begin{align}
    \ddot{X}_g - 4i\Omega \dot{X}_g + (\Delta_{\rm eff}^2 - 2i\dot{\Omega})X_g = \omega_g^2 V_g,\label{eqn:toy_eom2}
\end{align}
where $\Delta_{\rm eff}^2$ is the \emph{effective detuning}
\begin{align}
    \Delta_{\rm eff}^2 \equiv \omega_{\rm eff}^2 - 4\Omega^2.\label{eqn:det_def}
\end{align}
We can obtain the zeroth-order equilibrium solution to $X_g$ by ignoring the time derivatives in Eq.~(\ref{eqn:toy_eom2})
\begin{align}
    X_g\approx \frac{\omega_g^2 V_g}{\Delta_{\rm eff}^2}. \label{eqn:approx_sol0}
\end{align}
We can iteratively improve this solution by evaluating its time-derivative as
\begin{align}
    \dot{X}_g\approx-\bigg(\frac{1}{\Delta_{\rm eff}^2}\frac{\rm d}{{\rm d}t}\Delta_{\rm eff}^2 +\frac{3\dot{D}}{D}\bigg)X_g,
\end{align}
plugging the above expression back into Eq.~(\ref{eqn:toy_eom2}) and ignoring $\ddot{X}_g$, which yields \cite{lai94, yu24, nlrl_prl}
\begin{align}
    X_g \approx\frac{\omega_g^2V_g}{\Delta_{\rm eff}^2 - 2i\Omega\left[\frac{\dot{\Omega}}{\Omega} - \frac{2}{\Delta_{\rm eff}^2}\frac{{\rm d}\Delta_{\rm eff}^2}{{\rm d}t} - 6\frac{\dot{D}}{D}\right]}.\label{eqn:approx_sol}
\end{align}
Eq.~(\ref{eqn:approx_sol}) has two main implications: 
First, if the $l=|m|=2$ $g$-modes are linearly driven and evolve linearly, \emph{i.e.}, the effective frequency $\omega_{\rm eff}$ in the detuning in Eq.~\eqref{eqn:det_def} is the `bare' value $\omega_g$, they are resonantly amplified as the tidal forcing frequency ($2\Omega$) chirps through the eigenfrequency $\omega_g$, \emph{i.e.}, when $\omega_g^2 - 4\Omega^2$ changes sign. 
In fact, the equilibrium solution (\emph{i.e.}, Eq.~(\ref{eqn:approx_sol}) with $\Delta_{\rm eff}^2$ replaced with $\omega_g^2-4\Omega^2$) diverges at resonance, so we cannot obtain the amplitude of the linearly driven $g$-mode right at $2\Omega=\omega_g$ using this solution (we will address this issue shortly). 
Second, the nonlinear modification that takes $\omega_g \rightarrow \omega_{\rm eff}$ can significantly affect the dynamics as it changes the detuning parameter (potentially even dynamically, as we will see). 

Let us revisit how to calculate the amplitude of the \emph{linear} $g$-mode through resonance, following the approach in Lai~\cite{lai94c}. 
As mentioned earlier, we cannot use the equilibrium solution of Eq.~\eqref{eqn:approx_sol} (which assumes that the modes oscillate in phase with the driving). 
Instead, as the modes are excited and evolve through resonance, they end up oscillating at their natural frequency $\omega_g$, while the tidal potential drives the mode at $2\Omega$. 
Thus we have to go back to the equation of motion in Eq.~(\ref{eqn:toy_eom}) with $\omega_{\rm eff}^2\rightarrow \omega_g^2$, and define $\chi_g^{\rm (lin)}=\chi_g e^{2i\Phi+i\omega_g t}$. 
The rapidly varying phases of the modes and the orbit coherently add only when $2\Omega \approx \omega_g$. 
Thus, we can solve for the evolution of the amplitude $\chi_g^{\rm (lin)}$ using the stationary phase approximation, and obtain the amplitude at resonance, \emph{i.e.}, at $2\Omega=\omega_g$:
\begin{align}
|\chi_g^{\rm (lin)}|\approx \frac{1}{2}\sqrt{\frac{\pi}{4}}\frac{\omega_g}{\dot{\Omega}^{1/2}}|U_g|\bigg|_{2\Omega=\omega_g}\approx 10^{-4}.\label{eqn:lin_res_amp}
\end{align}
Here, the factor of $1/2$ on the right-hand side accounts for the fact that $|\chi_g^{\rm (lin)}|$ at $2\Omega=\omega_g$ is half of that at $\Omega\rightarrow \infty$. 

The nonlinear interaction we incorporate is initially small, and hence we can still compute the amplitude of the $g$-mode following similar steps taken in the linear theory. 
When the $g$-mode is driven far from resonance ($2\Omega\ll \omega_g$), its amplitude as given by Eq.~\eqref{eqn:approx_sol} is small, and grows approximately linearly as
\begin{align}
    |X_g|\propto |V_g| \propto \Omega^2. 
\end{align}
As a result, the nonlinear correction to the $g$-mode frequency $\omega_g$ evolves as
\begin{align}
    -\omega_g^2\kappa_{\rm eff}|X_g|^2\propto \Omega^4. \label{eq:omega4}
\end{align}
When the tidal forcing frequency $2 \Omega$ approaches $\omega_g$ from below, the $g$-mode is driven nearly resonantly, and its amplitude can become large enough for the nonlinear frequency shift to be significant. 
As Eq.~(\ref{eqn:toy_eff}) suggests in concert with Eq.~\eqref{eq:omega4}, if $\kappa_{\rm eff} < 0$ and the nonlinear corrections are subdominant, the $g$-mode's effective frequency $\omega_{\rm eff}$ shifts up faster~(as the fourth power of $\Omega$) than the rate at which tidal forcing frequency $2\Omega$ is itself evolving (which is, of course, linear in $\Omega$). 
This temporarily extends the duration for which the effective detuning of Eq.~\eqref{eqn:det_def} satisfies
\begin{align}
    \Delta_{\rm eff}^2 \gtrsim 0, \, \text{\emph{i.e.},}
\end{align}
prevents it from changing the sign immediately. 

\begin{figure}
    \centering
    \includegraphics[width=0.48\textwidth]{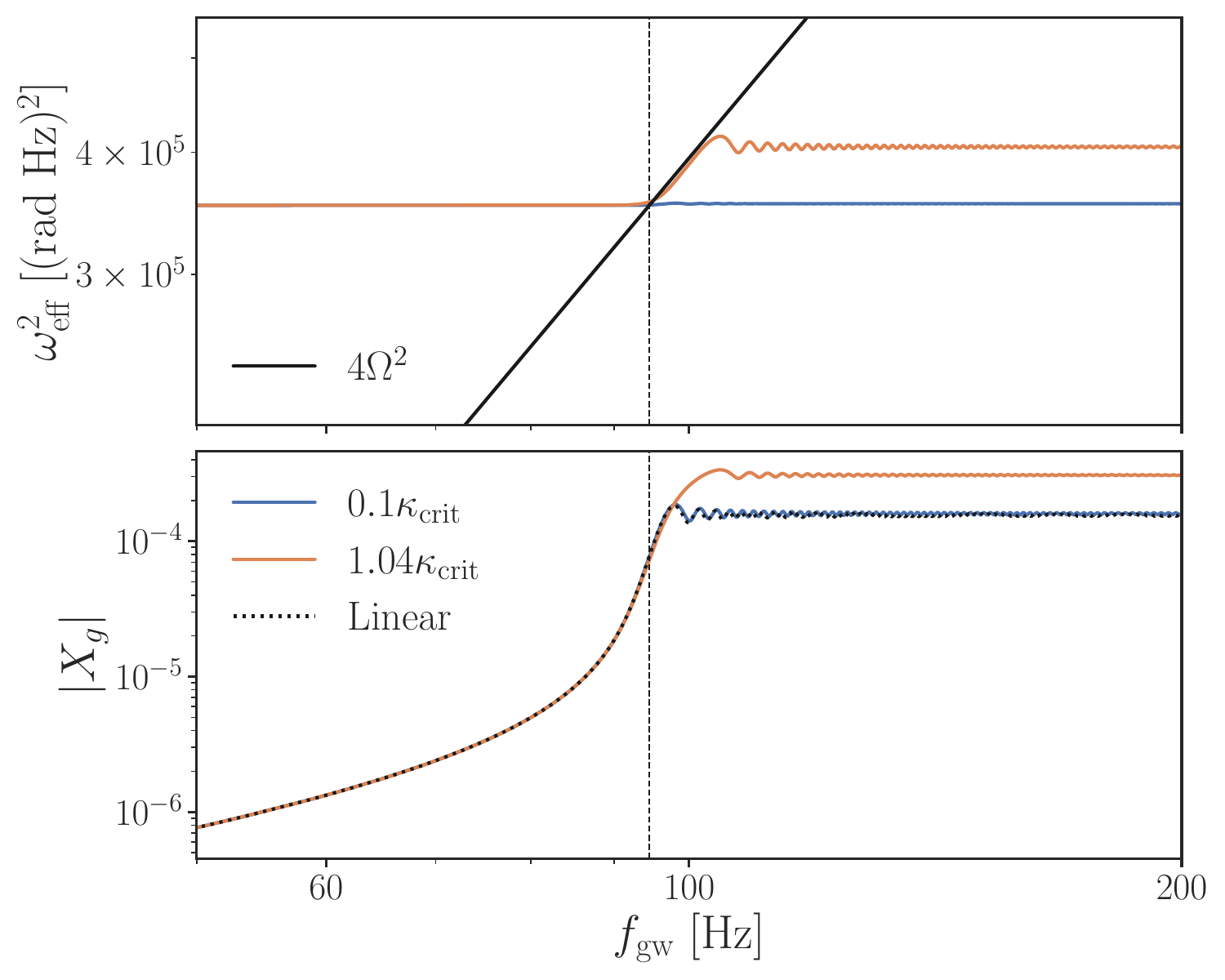}
    \caption{Demonstration of the lock initiating condition, as summarized in Eq.~(\ref{eqn:kap_crit}). \textit{Top}---The evolution of the effective frequencies of two systems with $\kappa_{\rm eff}=0.1\kappa_{\rm crit}$~(blue) and $1.04\kappa_{\rm crit}$~(orange). The effective (anti-) damping coefficient $\gamma_{\rm eff}$ is set to 0 for both systems. The vertical dashed line marks the linear resonance point, where $2\pi f_{\rm gw}=2\Omega=\omega_g$. The effective frequency briefly stays locked for $\kappa_{\rm eff}=1.04\kappa_{\rm crit}$. \textit{Bottom}---The amplitudes $X_g=\chi_ge^{-2i\Phi}$ of $l=m=2$ $g$-mode for the two systems. We additionally show the amplitude of the linearly driven $g$-mode~(black dotted). The system with $|\kappa_{\rm eff}|\ll |\kappa_{\rm crit}|$ behaves closely to the linear system.\label{fig:kap_crit}}
\end{figure}

For small values of $|\kappa_{\rm eff}|$, this leads to a negligible change in the mode's overall behavior, as we later show in Fig.~\ref{fig:kap_crit}.
However, if the effective coupling $|\kappa_{\rm eff}|$ is large enough, the phenomenology changes dramatically: the system can enter a locked state in which the detuning $\Delta_{\rm eff}^2$ is set to a small value and satisfies
\begin{align}
    \bigg|\frac{{\rm d}\Delta_{\rm eff}^2}{{\rm d}t}\bigg|\ll\frac{{\rm d}(4\Omega^2)}{{\rm d}t}\label{eqn:lock_def}
\end{align}
In other words, looking at the two terms in Eq.~\eqref{eqn:det_def}, the rate of change of the effective frequency $\dot{\omega}_{\rm eff}$ approximately matches the rate at which the tidal forcing changes:
\begin{equation}
    \frac{{\rm d}\omega^2_{\rm eff}}{{\rm d}t} = -\kappa_{\rm eff} \omega_{g}^2\frac{{\rm d}|X_g|^2}{{\rm d}t} \approx \frac{\rm d}{{\rm d}t}(4 \Omega^2).\label{eqn:lock_def2}
\end{equation}
Once locked, Eq.~(\ref{eqn:lock_def2}) determines the mode amplitude through:
\begin{align}
    |X_g|^2 &\approx \frac{\omega_g^2-4\Omega^2}{\omega_g^2\kappa_{\rm eff}} + C,\label{eqn:mode_amp0}
\end{align}
for $2\Omega \gtrsim \omega_g$, where $C$ is a constant of integration. 
The constant $C$ is the value of $|X_g|^2$ at $2\Omega=\omega_g$, which we can obtain from Eq.~(\ref{eqn:toy_eom2}).
Neglecting all the time derivatives therein, at the linear resonance point, \emph{i.e}, $2\Omega=\omega_g$, we can estimate
\begin{align}
    -\omega_g^2\kappa_{\rm eff}|X_g|^2X_g \approx \omega_g^2 V_g|_{2\Omega=\omega_g}.
\end{align}
Since $\kappa_{\rm eff}<0$, we write
\begin{align}
    C= \bigg(\frac{V_g|_{2\Omega=\omega_g}}{\kappa_{\rm eff}}\bigg)^{2/3}=\bigg[\frac{I_gW_{22}(R/D)^3|_{2\Omega=\omega_g}}{\kappa_{\rm eff}}\bigg]^{2/3},
\end{align}
where $I_g$ is the tidal overlap of the $g$-mode. We can plug this back into Eq.~(\ref{eqn:mode_amp0}) to get
\begin{align}
    |X_g|^2&\approx \frac{\omega_g^2-4\Omega^2}{\omega_g^2\kappa_{\rm eff}} + \bigg(\frac{W_{22}I_gR^3\Omega^2}{\kappa_{\rm eff}GM_{\rm t}}\bigg)^{2/3}\bigg|_{2\Omega=\omega_g}, \label{eqn:locked_amp}
\end{align}
where $M_{\rm t}$ is the binary's total mass we have used Kepler's law to eliminate $D$ in favor of $\Omega$.

We can find a constraint on the value of $\kappa_{\rm eff}$ for the locked state to be achieved as follows: the amplitude of the $g$-mode in the linear (thus non-locked) case evaluated at the linear resonance point at $2\Omega=\omega_g$, [given by Eq.~(\ref{eqn:lin_res_amp}]) has to be at least as large as the $g$-mode amplitude in Eq.~(\ref{eqn:locked_amp}) derived assuming a locked state; else the system never gets to the required level of nonlinearity for locking to happen. 
The constraint evaluates to 
\begin{align}
    |\kappa_{\rm eff}|\gg 
    \frac{10}{3\pi^{5/2}}\bigg(\frac{GM_{\rm t}}{I_g\Omega^2R^3}\bigg)^2\left(\frac{8\dot{\Omega}}{\omega_g^2}\right)^{3/2}\bigg|_{2\Omega=\omega_g}\equiv -\kappa_{\rm crit},
    \label{eqn:kap_crit}
\end{align}
where we plug the value of the coefficient $U_g$ from Eq.~\eqref{eq:Ug} into the linear amplitude in Eq.~\eqref{eqn:lin_res_amp}. In numerical estimates, we evaluate $\dot{\Omega}$ using the quadrupole formula:
\begin{align}
    \dot{\Omega}=\frac{96G^{5/3}}{5c^5}\eta M_{\rm t}^{5/3}\Omega^{11/3}. \label{eq:quadrupole}
\end{align}
This relation was also presented in \citetalias{nlrl_prl} [see Eq.~(11) therein].
We show the value of $\kappa_{\rm crit}$ for $n=1$, $l=|m|=2$ $g$-modes in Table~\ref{tab:eff_vals}.

\begin{figure}
    \centering
    \includegraphics[width=0.48\textwidth]{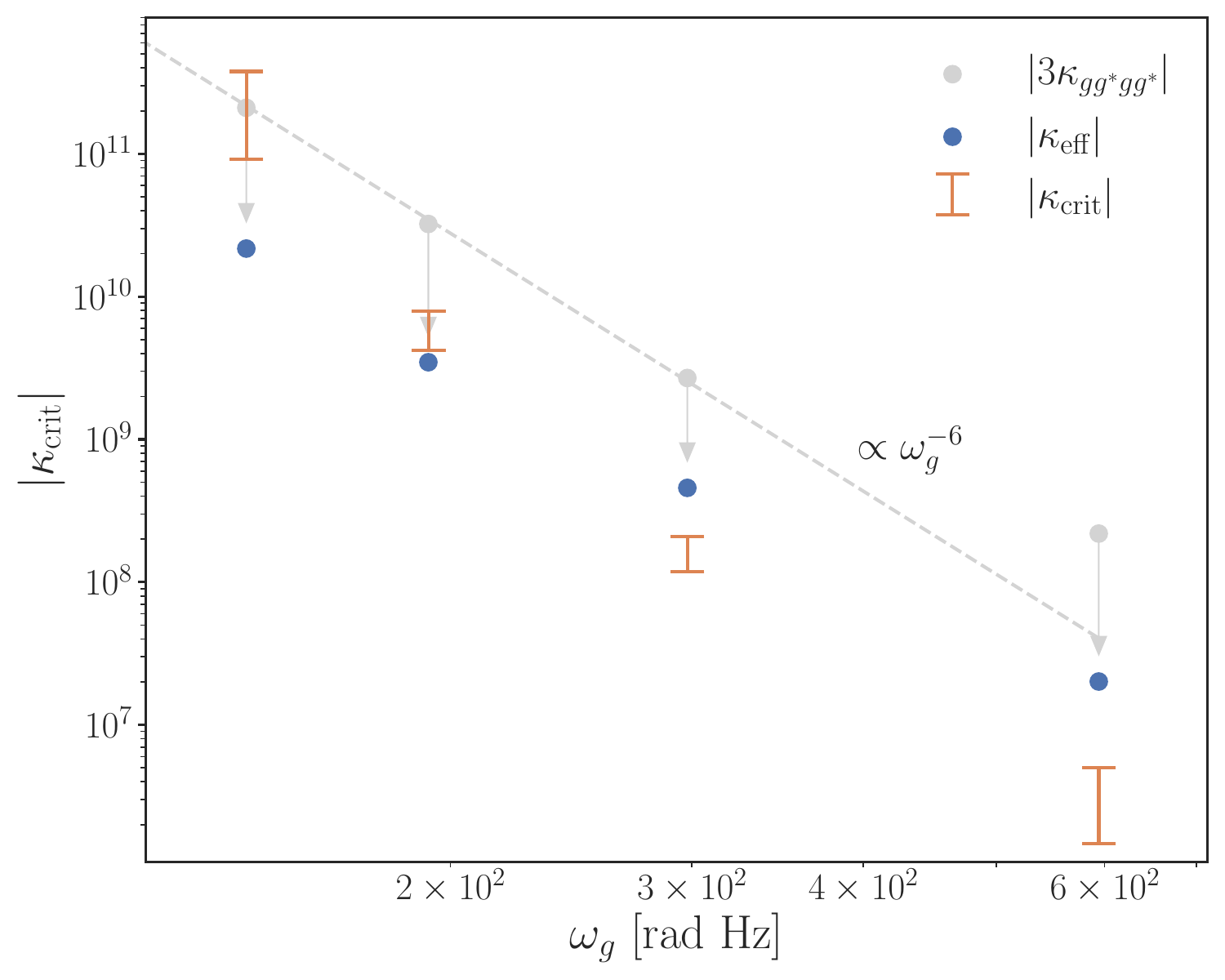}
    \caption{Comparison between $\kappa_{\rm crit}$~(orange) and $\kappa_{\rm eff}$~(blue) for the $1\leq n\leq 4$, $l=2$ $g$-modes. We calculate $\kappa_{\rm crit}$ using Eq.~(\ref{eqn:kap_crit}) for a $1.4$-$1.4\,M_{\odot}$ NS binary, from which we expect $\kappa_{\rm crit}\propto \omega_g^{-6.5}$. We additionally show the values of $3\kappa_{gg^\ast gg^\ast}$~(gray circles), as well as the effect of cancellation due to adiabatically sourced modes~(gray arrows). 
    Only the lowest order ($n=1$, highest frequency) $g$-mode satisfies $|\kappa_{\rm eff}| \gg |\kappa_{\rm crit}|$. We expect the lock will not initiate for the $n=3$, $4$ $g$-modes they do not satisfy $|\kappa_{\rm eff}|\gg |\kappa_{\rm crit}|$.
    }
    \label{fig:kap_crit2}
\end{figure}

In Fig.~\ref{fig:kap_crit}, we demonstrate the above condition by evolving Eq.~(\ref{eqn:toy_eom}) for some representative values of the effective coupling coefficients. 
We show two cases with $\kappa_{\rm eff}=1.04 \, \kappa_{\rm crit}$~(orange) and $0.1 \, \kappa_{\rm crit}$~(blue), and set $\gamma_{\rm eff}$ to 0 for both systems. 
In the bottom panel, we show the mode amplitude $|X_g|$. The system with $\kappa_{\rm eff}=0.1\kappa_{\rm crit}$ behaves very closely to the linearly driven $l=m=2$ $g$-mode~(black dotted line), while the system with $\kappa_{\rm eff}=1.04\kappa_{\rm crit}$ has an enhanced amplitude past the linear resonance point~(vertical dashed line). 
The top panel shows the effective frequency $\omega_{\rm eff}$ (which is a function of the mode amplitude): past the linear resonance point, it is locked to the driving frequency for a short duration for $\kappa_{\rm eff}=1.04 \, \kappa_{\rm crit}$. We will see later that including the $\gamma_{\rm eff}$ coefficient enables the lock to be maintained for much longer.  

In Fig.~\ref{fig:kap_crit2}, we compare $\kappa_{\rm crit}$~[orange, computed using Eq.~\eqref{eqn:kap_crit}] and $\kappa_{\rm eff}$~(blue) for $1\leq n \leq 4$, $l=2$, $g$-modes.
We start by showing the `bare' value of $3 \, \kappa_{gg^\ast gg^\ast}$~(light gray markers). 
Then, we show the cancellation due to the terms involving adiabatically sourced modes (extra terms in Eq.~\eqref{eqn:k_eff_0}) using arrows that lead from the sizes of the bare terms to the final value of $\vert \kappa_{\rm eff} \vert$. 
Only the lowest-order $g$-mode satisfies $|\kappa_{\rm eff}|\approx 2\times {10^7}\gg |\kappa_{\rm crit}|$. 
We can also expect that the lock will not initiate for the $n=3$ and $n=4$ $g$-modes since they do not satisfy  $|\kappa_{\rm eff}|\gg |\kappa_{\rm crit}|$. 
We can also understand the dominant scaling of the critical and the bare four-mode coupling terms with the mode frequency. 
We additionally note that since $\dot{\Omega}\propto \Omega^{11/3}$ [from the quadrupole formula, Eq.~\eqref{eq:quadrupole}] and it was numerically found in Yu~\emph{et~al.} \cite{yu17} that $I_g\sim \omega_{g}^{2.5}$.
Then, we can estimate 
\begin{align}
    \kappa_{\rm crit}\sim \frac{\dot{\Omega}^{3/2}}{I_g^2\Omega^4 \omega_g^3}\bigg|_{2\Omega=\omega_g}\sim \omega_g^{-6.5}.
\end{align}
Meanwhile, the Wentzel-Kramers-Brillouin~(WKB) analysis yields that the high-order $g$-modes asymptotically follow the scaling $\xi_g^{\rm h}\sim \omega_g^{-1}$ and ${\rm d}\xi_g^{\rm h}/{\rm d}r\sim \omega_g^{-2}$ \cite{aerts10}, which suggests
\begin{align}
    \kappa_{gg^\ast gg^\ast}\sim \bigg[\frac{1}{r}\frac{{\rm d}(\xi_g^{\rm h} \xi_{g^\ast}^{\rm h})}{{\rm d}r}\bigg]^2\sim \omega_g^{-6}.
\end{align}
where we have used the fact that $\kappa_{gg^\ast gg^\ast}$ is dominated by $[h^{(2)}_{gg^\ast}]^2E^{(22)}_{gg^\ast,gg^\ast}$ in Eq.~(158) of Weinberg~\cite{w16}.

Similarly, in the three-mode coupling, $\kappa_{gg\beta}\sim \omega_{\beta}^2 \xi_{\beta}^{\rm r}\xi_{g}^{\rm h}\xi_{g}^{\rm h} \sim \omega_\beta \omega_g^{-2}\sim \omega_g^{-3}$ [\emph{cf.} Eq.~(A60) of Weinberg \emph{et~al.}~\cite{w12}], where the last equality follows from matching the spatial wavenumbers $k$ of the $p$- and $g$-modes which have the dispersion relations
\begin{align}
    k_p\sim \omega_p&&{\rm and }&& k_g\sim \omega_g^{-1}.
\end{align}
When the three-mode coupling is chained to form an effective four-mode coupling, we also have $\kappa_{gg\beta}^2\sim\omega_g^{-6}$ (which is understandable as it cancels the `bare' four-mode coupling term). 
Hence, the critical value $\kappa_{\rm crit}$ rises in a steeper manner than the four-mode coupling term $\kappa_{gg^\ast gg^\ast}$ when $\omega_g$ decreases with the increasing order of the $g$-mode. 
This makes the lowest-order $g$-mode the most likely mode to enter RL.

\subsubsection{Maintaining the lock\label{subsubsec:maintain}}
So far, we have only considered the effect of the $\kappa_{\rm eff}$ coefficient, which modifies the real part of the mode's frequency. 
If we consider Eq.~(\ref{eqn:approx_sol}), in the denominator, the term proportional to $2i\Omega$ corresponds to the damping induced by the inspiral (not to be confused with the GW-damping directly acting on modes, which we ignore in this paper). 
In the results presented in Fig.~\ref{fig:kap_crit}, this inspiral damping rapidly broadens the width of the Lorentzian multiplying $V_g$ in Eq.~(\ref{eqn:approx_sol}), and eventually breaks the lock. 
If the frequency shift in $\omega_{\rm eff}^2$ (and hence $\Delta^2_{\rm eff}$) has an imaginary part, it modifies the rate at which the mode damps. 
For $n=1$, $l=|m|=2$ $g$-modes, we find that the imaginary shift cancels the inspiral damping and keeps the Lorentzian's width nearly constant and leaves the $g$-modes locked in a near-resonant state throughout the merger (see Fig.~3 in \citetalias{nlrl_prl}).

To get a quantitative estimate, we begin by noting that if the mode stays locked well after resonance, we can neglect the constant terms in Eq.~(\ref{eqn:locked_amp}), which leads to the frequency scaling for the amplitude
\begin{align}
    &|X_g|^2\sim {\rm Re}[X_g]^2\propto \Omega^2, &\text{since}&&\kappa_{\rm eff}<0.\label{eqn:real_scaling}
\end{align}
Plugging the scaling of Eq.~\eqref{eqn:real_scaling} into the solution in Eq.~(\ref{eqn:approx_sol}), and noting that the driving $V_g\propto \Omega^2$, we get that the detuning $\Delta_{\rm eff}^2\propto \Omega$ approximately before the terms within the parentheses in the denominator become significant.
Then, with 
\begin{align}
    &\frac{1}{\Delta_{\rm eff}^2}\frac{{\rm d}\Delta_{\rm eff}^2}{{\rm d}t}\approx \frac{\dot{\Omega}}{\Omega},
\end{align}
and Kepler's third law, 
\begin{align}
\frac{\dot{D}}{D}= -\frac{2\dot{\Omega}}{3\Omega},
\end{align}
we can rewrite the solution in Eq.~(\ref{eqn:approx_sol}) as 
\begin{align}
    X_g &\approx \frac{\omega_g^2 V_g}{\Delta_{\rm eff}^2 - 6i\dot{\Omega}}\nonumber \\
    &=\frac{\omega_g^2 V_g}{{\rm Re}[\Delta_{\rm eff}^2] + i \, {\rm Im}[\Delta_{\rm eff}^2] - 6i\dot{\Omega}}\nonumber\nonumber \\
    &\approx \frac{\omega_g^2 V_g}{{\rm Re}[\Delta_{\rm eff}^2]}\bigg(1+i\frac{6\dot{\Omega} - {\rm Im}[\Delta_{\rm eff}^2]}{{\rm Re}[\Delta_{\rm eff}^2]}\bigg). \label{eqn:lock_sol}
\end{align}
This assumes that the imaginary terms in the solution for $X_g$ are small compared to the real-valued ones. 
Examining the last line, we find the condition for the lock to be sustained is
\begin{align}
    \bigg|\frac{6\dot{\Omega}-{\rm Im}[\Delta_{\rm eff}^2]}{{\rm Re}[\Delta_{\rm eff}^2]}\bigg|\ll 1.\label{eqn:break_cond}
\end{align}
Otherwise, $X_g$ does not scale as $\Omega$, and the RL condition in Eq.~(\ref{eqn:lock_def}) no longer holds. 

From this, we can derive a critical anti-damping coefficient $\gamma_{\rm crit}$ required to cancel the damping for a significant portion of the inspiral.
As the mode is entering the locked phase, ${\rm Im}[\Delta_{\rm eff}^2]$ is initially small compared to $6\dot{\Omega}$. 
Therefore, we can use Eq.~(\ref{eqn:lock_sol}) to approximately write ${\rm Im}[X_g]$ as
\begin{align}
    {\rm Im}[X_g]&\approx\frac{6\omega_g^2 V_g \dot{\Omega}}{[{\rm Re}[\Delta_{\rm eff}^2]]^2}\propto \Omega^{11/3}.\label{eqn:im_scaling_no_canc}
\end{align}
This leads to 
\begin{align}
    {\rm Im}[\Delta_{\rm eff}^2]=\omega_g^2\gamma_{\rm eff}{\rm Im}[X_fX_g^\ast]\propto \Omega^{17/3},
\end{align}
where we have used $X_f\approx {\rm Re}[X_f]\propto \Omega^2$.
Therefore, the ${\rm Im}[\Delta_{\rm eff}^2]$ term in Eq.~\eqref{eqn:lock_sol} initially starts out smaller than the $6\dot{\Omega}$ term, but the former grows as $\Omega^{17/3}$ while the latter $6\dot{\Omega} \propto \Omega^{11/3}$, so eventually the two terms can reach a similar size. 
At this point, the mode can reach a state where the numerator in Eq.~(\ref{eqn:break_cond}) vanishes
\begin{align}
{\rm Im}[\Delta_{\rm eff}^2]=\omega_g^2\gamma_{\rm eff}{\rm Im}[X_f X_g^\ast]=6\dot{\Omega},\label{eqn:canc}
\end{align}
if $\gamma_{\rm eff}I_f I_g >0$. 
If the cancellation in Eq.~(\ref{eqn:canc}) is achieved during the locked phase, the imaginary part of the mode amplitude ${\rm Im}[X_g]$ then dynamically keeps adjusting itself during the rest of the inspiral to maintain the cancellation (\emph{i.e.}, cancel the effect of the GW damping) and maintains the lock. 
Since $6\dot{\Omega}\propto \Omega^{11/3}$ and $X_f\approx {\rm Re}[X_f]\propto \Omega^2$, we can estimate from Eq.~(\ref{eqn:canc}) that the imaginary component of the mode is locked and scales with frequency as
\begin{align}
    {\rm Im}[X_g]\propto \Omega^{5/3},\label{eqn:im_scaling}
\end{align}
which we will verify shortly.

\begin{figure}[t]
    \centering
    \includegraphics[width=0.48\textwidth]{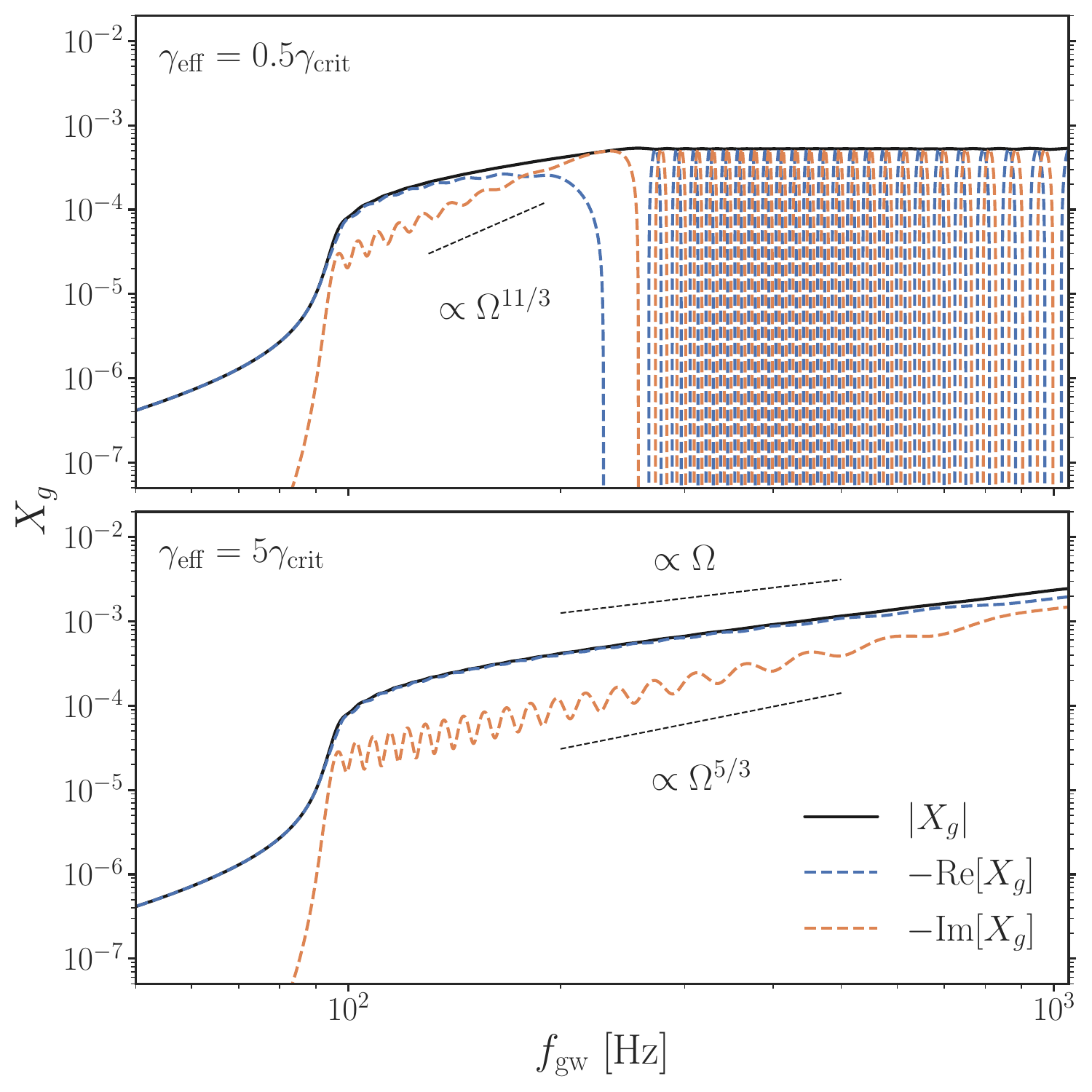}
    \caption{Amplitude of $n=1$, $l=m=2$  $g$-mode computed using Eq.~(\ref{eqn:toy_eom2}) with $\kappa_{\rm eff}\approx -2\times 10^7$, and $\gamma_{\rm eff}=0.5\gamma_{\rm crit}$ (top) and $\gamma_{\rm eff}=5\gamma_{\rm crit}$ (bottom). 
    We show the magnitudes (black), real (blue), and imaginary parts (orange) of the amplitudes.
    \emph{Top}---Since $|\gamma_{\rm eff}|<|\gamma_{\rm crit}|$, ${\rm Im}[\Delta_{\rm eff}^2]$ does not grow comparable to $6\dot{\Omega}$ and the system follows Eq.~(\ref{eqn:im_scaling_no_canc}).
    \emph{Bottom}---The lock is achieved for both the real and imaginary amplitudes of the $g$-modes, and we can verify Eq.~(\ref{eqn:im_scaling}).}
    \label{fig:gam_crit}
\end{figure}

Whether the imaginary component can reach the locked state or not depends on the value of $\gamma_{\rm eff}$.
In Eq.~\eqref{eqn:lock_sol}, without the presence of ${\rm Im}[\Delta_{\rm eff}^2]$, the inspiral damping breaks the lock when $6\dot{\Omega}\sim {\rm Re}[\Delta_{\rm eff}^2]$.
Hence, ${\rm Im[\Delta_{\rm eff}^2]}$ must cancel the inspiral damping, at latest, when $6\dot{\Omega}\sim {\rm Re}[\Delta_{\rm eff}^2]$.
This defines the critical value of the anti-damping coefficient
\begin{align}
    \gamma_{\rm crit}\equiv \frac{6\dot{\Omega}}{\omega_g^2 {\rm Im}[X_fX_g^\ast]}\bigg|_{{\rm Re}[\Delta_{\rm eff}^2]=6\dot{\Omega}}.\label{eqn:gamcrit0}
\end{align}
To find the point where ${\rm Re}[\Delta_{\rm eff}^2]=6\dot{\Omega}$, we can use Eqs.~(\ref{eqn:approx_sol}) and (\ref{eqn:locked_amp}) to write an alternate expression for $\Delta_{\rm eff}^2$
\begin{align}
    \Delta_{\rm eff}^2&\approx \frac{\omega_g^2 V_g}{X_g}\approx \frac{\omega_g^3V_g|\kappa_{\rm eff}|^{1/2}}{2\Omega},\label{eqn:det2}
\end{align}
where the second equality holds for $2\Omega \gg \omega_g$.
Using the above expression with the quadrupole formula, we can rewrite ${\rm Re}[\Delta_{\rm eff}^2]=6\dot{\Omega}$ as
\begin{align}
    \frac{|\kappa_{\rm eff}|^{1/2}|I_g|W_{22}\omega_g^3 R^3}{2GM_{\rm t}}\Omega=\frac{576G^{5/3}}{5c^5}\eta M_{\rm t}^{5/3}\Omega^{11/3},
\end{align}
Using $\Omega$ found above, we can evaluate Eq.~\eqref{eqn:gamcrit0} using Eq.~\eqref{eqn:im_scaling_no_canc} for ${\rm Im}[X_g]$, quadrupole formula for $6\dot{\Omega}$, and $X_f\approx {\rm Re}[X_f]\approx V_f$.
Then, we find the condition for the lock to persist for a significant portion of the inspiral
\begin{align}
    &|\gamma_{\rm eff}|\gtrsim |\gamma_{\rm crit}|, \, {\rm where} \label{eqn:gam_cond}\\
    &\gamma_{\rm crit}=24\bigg(\frac{96|\kappa_{\rm eff}|^5\eta^6 I_g^2}{125\pi^3I_f^8}\bigg)^{1/8} \bigg(\frac{GM_{\rm t}}{R^3\omega_g^2}\bigg)^2\bigg(\frac{R\omega_g}{c}\bigg)^{15/4}.\label{eqn:gam_crit}
\end{align}
Table~\ref{tab:eff_vals} shows the value of $\gamma_{\rm crit}$ for the $n=1$, $l=|m|=2$ $g$-modes.

In Fig.~\ref{fig:gam_crit}, 
we show the $n=1$, $l=m=2$ $g$-mode amplitude evolved using Eq.~(\ref{eqn:toy_eom2}) with $\kappa_{\rm eff}\approx -2\times 10^7$ and $\gamma_{\rm eff}=0.5 \, \gamma_{\rm crit}$~(top) and $5 \, \gamma_{\rm crit}$~(bottom). 
We show the magnitudes~(black), real~(blue), and imaginary parts~(orange) of the amplitude.
In the top panel where $|\gamma_{\rm eff}|  <|\gamma_{\rm crit}|$, the cancellation ${\rm Im}[\Delta_{\rm eff}^2]$ between $6\dot{\Omega}$ is negligible, and the imaginary part of the amplitude ${\rm Im}[X_g]$ is dominantly sourced according to Eq.~(\ref{eqn:im_scaling_no_canc}).
The lock breaks shortly after initiation due to the inspiral damping. 
In the bottom panel where $|\gamma_{\rm eff}|>|\gamma_{\rm crit}|$, the system evolves into a locked state in which we can verify Eqs.~(\ref{eqn:real_scaling}) and (\ref{eqn:im_scaling}). 

\begin{figure}
    \centering
    \includegraphics[width=0.48\textwidth]{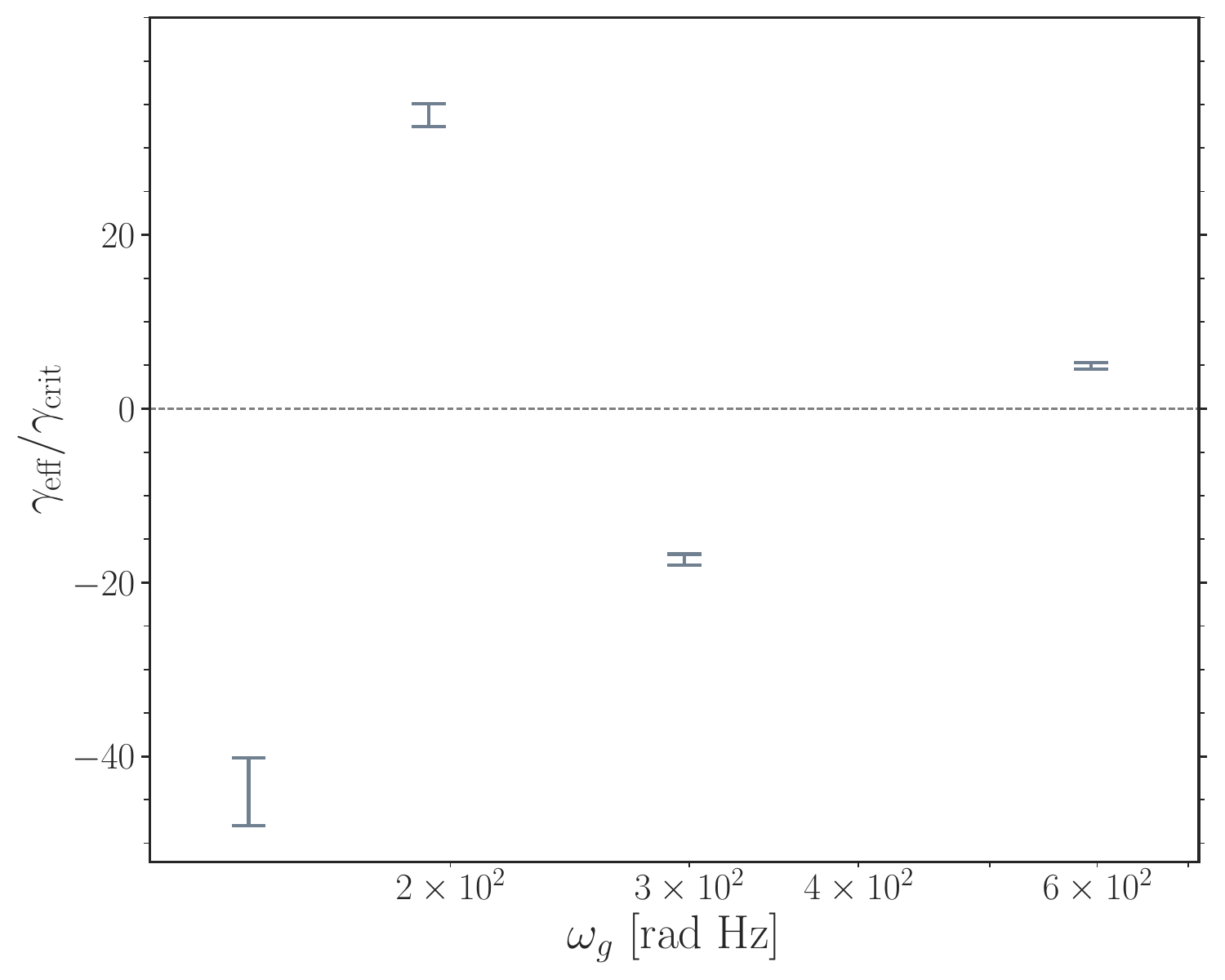}
    \caption{Ratios $\gamma_{\rm eff}/\gamma_{\rm crit}$ for the $1\leq n \leq 4$, $l=2$ $g$-modes. 
    The values of $\gamma_{\rm crit}$ are calculated for$1.4$-$1.4\, M_\odot$ NS binary and their signs satisfy $\gamma_{\rm crit} I_{g_n} I_f > 0$.
    As $\gamma_{\rm eff}I_{g_n}I_f<0$ for $n=2$ and $4$, we expect that lock will break shortly after initiation, although the latter will not enter the locked phase at all since $|\kappa_{\rm eff} < |\kappa_{\rm crit}|$. 
    For $n=3$, $\gamma_{\rm eff}\approx 30 \gamma_{\rm crit}$, although the lock does not initiate as $|\kappa_{\rm eff}| < |\kappa_{\rm crit}|$ as well.
    We have numerically verified that for $2 \leq n \leq  4$, $l=|m|=2$ $g$-modes, RL is negligible.
    }
    \label{fig:gam_crit2}
\end{figure}

We also compare $\gamma_{\rm eff}$ with $\gamma_{\rm crit}$ for $1 \leq n \leq 4$, $l=2$ $g$-modes in Fig.~\ref{fig:gam_crit2}. 
We show the ratios $\gamma_{\rm eff}/\gamma_{\rm crit}$ calculated for $1.4$-$1.4\, M_\odot$ NS binary where the sign of $\gamma_{\rm crit}$ satisfies $\gamma_{\rm crit}I_{g_n}I_f>0$.
For the $n=1$, $l=|m|=2$ $g$-modes, we find $\gamma_{\rm eff} I_f I_g>0$, and $\gamma_{\rm eff}\approx 5\gamma_{\rm crit}$. 
Recalling Fig.~\ref{fig:kap_crit2}, this mode satisfies $|\kappa_{\rm eff}|\gg |\kappa_{\rm crit}|$, hence we expect that the mode enters the locked state and stays in it until the late stage of the inspiral, which we also numerically verify in Sec.~\ref{subsec:num} (see Fig.~\ref{fig:full_sol}). 
As for the other modes, the $n=2$, $l=|m|=2$ $g$-modes satisfy $\gamma_{\rm eff}I_fI_g <0$ and hence the lock if established will break immediately. 
The $n=3$, $n=4$ $g$-modes never enter the locked phase as they do not satisfy $|\kappa_{\rm eff}|\gg |\kappa_{\rm crit}|$ (see Fig.~\ref{fig:kap_crit2}), so their value of $\gamma_{\rm eff}$ is immaterial. 
We have numerically verified that RL is negligible for the $2\leq n\leq 4$, $l=|m|=2$ $g$-modes, and hence conclude that RL is significant only for the lowest-order $g$-modes.

\subsection{Hamiltonian of the System\label{subsec:ham}}
In the previous section, we have directly resorted to using equations of motion to gain an overall understanding of the phenomenology of RL. 
In this section, we provide the full Hamiltonian of the NS from which we derive the equations of motion of modes.

We start with the intrinsic and external components of the Lagrangian in Sec.~\ref{sec:form}.
We construct the intrinsic Hamiltonian $H_{\rm int}$ from  Eq.~(\ref{eqn:l_int_3}), including terms up to fourth-order in mode amplitude. 
In our normalization, its generic expression is given as \cite{vhoolst94}
\begin{align}
    \frac{H_{\rm int}}{E_0}&=\frac{1}{2}\sum_{a}\bigg(\frac{|\dot{\chi}_a|^2}{\omega_a^2} + |\chi_a|^2 \bigg) - \frac{1}{3}\sum_{abc}\kappa_{abc}\chi_a^\ast \chi_b^\ast\chi_c^\ast \nonumber \\
    &\phantom{=}- \frac{1}{4}\sum_{abcd}\kappa_{abcd}\chi_a^\ast\chi_b^\ast\chi_c^\ast\chi_d^\ast,
\end{align}
where we use $\chi_{a^\ast}$ and $\dot{\chi}_{a}E_0/\omega_a^2$ as the generalized coordinates and momenta. Note that $\chi_{a^\ast}=\chi_a^\ast$ is the amplitude of the mode with that is the complex conjugate of $a$. 
 
We split the Hamiltonian as $H_{\rm int}=H_{\rm int}^{fg} + H_{\rm int}^\beta$. 
The first component depends on the amplitudes of the $l=2$ $f$- and $g$-modes, the only modes that we numerically integrate. 
Its expression reads
\begin{align}
    \frac{H^{fg}_{\rm int}}{E_0}&=\frac{1}{2}\sum_a^{\in\{f,g\}}\bigg(\frac{|\dot{\chi}_a|^2}{\omega_a^2} + |\chi_a|^2\bigg)-\frac{1}{3}\sum_{abc}^{\in\{f,g\}}\kappa_{abc}\chi_a^\ast \chi_b^\ast \chi_c^\ast\nonumber \\
    &\phantom{=}-\frac{1}{4}\sum_{abcd}^{\in\{f,g\}}\kappa_{abcd}\chi_a^\ast\chi_b^\ast\chi_c^\ast\chi_d^\ast.
\end{align}
The superscript ``$\in\{f,g\}$'' indicates the summation runs over the $l=2$ $f$- and $g$-modes. 

The second component $H_{\rm int}^\beta$ sums over the adiabatically sourced modes and is given by 
\begin{align}
    \frac{H^\beta_{\rm int}}{E_0}\approx \frac{1}{2}\sum_{\beta}|\chi_\beta|^2+\frac{H^{ff\beta}_{\rm int}}{E_0} +\frac{H^{fg\beta}_{\rm int}}{E_0} + \frac{H^{gg\beta}_{\rm int}}{E_0},
\end{align}
where we have ignored the associated kinetic terms $|\dot{\chi}_\beta|^2/\omega_\beta^2$.
The last three terms on the right-hand side describe the three-mode interaction energy between the adiabatically sourced modes, $l=2$ $f$- and $g$-modes.
Their expressions are 
\begin{widetext}
\begin{align}
    -\frac{H^{gg\beta}}{E_0}&=\sum_\beta \left[2\kappa_{gg\beta_{-4}}{\rm Re}[(\chi_{g}^\ast)^2 \chi_{\beta_{-4}}^\ast] + 4\kappa_{gg_0\beta_{-2}}{\rm Re}[\chi_{g}^\ast \chi_{g_0}^\ast \chi_{\beta_{-2}}^\ast]+ 2\kappa_{gg^\ast\beta_{0}}|\chi_g|^2 \chi_{\beta_{0}}^\ast
    + \kappa_{g_0g_{0}\beta_{0}}(\chi_{g_0}^\ast)^2  \chi_{\beta_{0}}^\ast\right],\label{eqn:pham_1}\\
    -\frac{H^{gf\beta}}{E_0}&=\sum_\beta\Big[ 4\kappa_{gf\beta_{-4}}{\rm Re}[\chi_{g}^\ast \chi_f^\ast \chi_{\beta_{-4}}^\ast] +4\kappa_{fg_0\beta_{-2}}{\rm Re}[\chi_f^\ast \chi_{g_0}^\ast \chi_{\beta_{-2}}^\ast] +4\kappa_{f_0g\beta_{-2}}{\rm Re}[\chi_{f_0}^\ast \chi_{g}^\ast \chi_{\beta_{-2}}^\ast] \nonumber \\
    \phantom{-\frac{H^{gf\beta}}{E_0}}&\phantom{=\sum_\beta} \ +4\kappa_{f g^\ast\beta_0} {\rm Re}[\chi_f^\ast \chi_g \chi_{\beta_0}^\ast]  +2\kappa_{f_{0} g_{0}\beta_0} \chi_{f_0}^\ast \chi_{g_0}^\ast \chi_{\beta_0}^\ast\Big],
    \label{eqn:pham_2}\\
    -\frac{H^{ff\beta}}{E_0}&=\sum_\beta \left[2\kappa_{ff\beta_{-4}}{\rm Re}[(\chi_{f}^\ast)^2 \chi_{\beta_{-4}}^\ast] + 4\kappa_{ff_0\beta_{-2}}{\rm Re}[\chi_{f}^\ast \chi_{f_0}^\ast \chi_{\beta_{-2}}^\ast]+ 2\kappa_{ff^\ast\beta_{0}}|\chi_f|^2 \chi_{\beta_{0}}^\ast
    +\kappa_{f_0f_{0}\beta_{0}}(\chi_{f_0}^\ast)^2  \chi_{\beta_{0}}^\ast\right],\label{eqn:pham_3}
\end{align}
\end{widetext}
where we have explicitly indicated the azimuthal quantum numbers of the modes using subscripts, except for $l=m=2$ $f$- ($\chi_f$) and $g$-modes ($\chi_g$).
Also, we have used $\chi_{\beta_{m_\beta}}^\ast=\chi_{\beta_{-m_\beta}}$ and $\kappa_{abc}=\kappa_{a^\ast b^\ast c^\ast}$ to compactly write $\kappa_{abc}\chi_a^\ast \chi_b^\ast \chi_c^\ast + \kappa_{a^\ast b^\ast c^\ast }\chi_{a^\ast}^\ast \chi_{b^\ast}^\ast \chi_{c^\ast}^\ast = 2\kappa_{abc}{\rm Re}[\chi_a^\ast \chi_b^\ast \chi_c^\ast]$. 

We similarly construct and split the external Hamiltonian as $H_{\rm ext}=H_{\rm ext}^{fg} + H_{\rm ext}^{\beta}$, where
\begin{align}
    &\frac{H_{\rm ext}^{fg}}{E_0}= -\sum_a^{\in\{f,g\}}U_a\chi_a^\ast -\frac{1}{2}\sum_{ab}^{\in\{f,g\}}U_{ab}\chi_a\chi_b, \\
    &\frac{H_{\rm ext}^{\beta}}{E_0}=-\sum_{\beta,m}(U_{\beta f_m}\chi_{\beta} \chi_{f_m}+U_{\beta g_m}\chi_{\beta} \chi_{g_m}).
\end{align}

Then, the total Hamiltonian of the NS is
\begin{align}
    H_\ast = H_{\rm int}^{fg} + H_{\rm int}^\beta + H_{\rm ext}^{fg} + H_{\rm ext}^\beta.
\end{align}
We derive the equations of motion of each mode using 
\begin{align}
    \ddot{\chi}_aE_0/\omega_a^2=-\partial H_\ast/\partial \chi_a^\ast.
\end{align}
Ignoring the left-hand side of the above equation, the expressions for the high-frequency, adiabatically sourced modes are
\begin{widetext}
\begin{align}
&\chi_{\beta_{4}} = \kappa_{\beta_{4}f^\ast f^\ast  } \chi_f^2 +\kappa_{\beta_{4}g^\ast g^\ast }\chi_g^2 + 2\kappa_{\beta_{4}f^\ast g^\ast} \chi_f \chi_g +  U_{\beta_{4}f^\ast}^\ast \chi_f
+ U_{\beta_{4}g^\ast }^\ast \chi_g ,\label{eqn:p5}\\
&\chi_{\beta_{2}} = 2\kappa_{\beta_{2}g^\ast g_0}\chi_g \chi_{g_0} + 2\kappa_{\beta_{2}f^\ast g_0}\chi_f \chi_{g_0} +2\kappa_{\beta_{2}f_0 g^\ast } \chi_{f_0} \chi_g + 2\kappa_{\beta_{2}f^\ast f_0} \chi_f \chi_{f_0}\nonumber\\
&\phantom{\chi_{\beta_{2}}=}+U_{\beta_{2}f_0}^\ast \chi_{f_0}^\ast+U_{\beta_2f^\ast}^\ast \chi_{f}+U_{\beta_{2}g_0}^\ast \chi_{g_0}^\ast+U_{\beta_{2}g^\ast}^\ast \chi_{g},\label{eqn:p6}\\
&\chi_{\beta_{0}} = 2(\kappa_{\beta_0g g^\ast }|\chi_g|^2+\kappa_{\beta_0f f^\ast}|\chi_f|^2 + \kappa_{\beta_0f g^\ast}\chi_f \chi_g^\ast + \kappa_{\beta_0f^\ast g}\chi_f^\ast \chi_g + \kappa_{\beta_0f_0g_0}\chi_{f_0} \chi_{g_0}) + \kappa_{\beta_0 g_0g_0 }\chi_{g_0}^2 + \kappa_{\beta_0f_0f_0}\chi_{f_0}^2 \nonumber\\
&\phantom{\chi_{\beta_0}=} + U_{\beta_0 g}^\ast \chi_g^\ast+ U_{\beta_0g^\ast}^\ast \chi_g + U_{\beta_0g_0}^\ast \chi_{g_0}^\ast + U_{\beta_0 f}^\ast \chi_f^\ast+ U_{\beta_0f^\ast}^\ast \chi_f  + U_{\beta_0f_0}^\ast \chi_{f_0}^\ast. \label{eqn:p7}
\end{align}
\end{widetext}
We obtain the equations of motion for $l=2$ $f$- and $g$-modes using a similar procedure. 
Below, we focus on the $l=m=2$ $g$-mode (we repeat the same procedure for the other modes).
We again obtain the equation of motion from the Hamiltonians
\begin{align}
    \ddot{\chi}_gE_0/\omega_g^2 
    &= -\frac{\partial (H^{fg}_{\rm int} + H^{fg}_{\rm ext})}{\partial \chi_g^\ast} -\frac{\partial (H^\beta_{\rm int} + H^{\beta}_{\rm ext})}{\partial \chi_g^\ast}.\label{eqn:g_eom_1}
\end{align}
The first term on the right-hand side evaluates to
\begin{multline}
   -\frac{1}{E_0}\frac{\partial(H^{fg}_{\rm int} + H^{fg}_{\rm ext})}{\partial \chi_g^\ast} =-\chi_g + U_g +\sum_{bc}^{\in\{f,g\}}\kappa_{gbc}\chi_b^\ast \chi_c^\ast  \\
   + \sum_{a}^{\in\{f,g\}}U_{ga}^\ast \chi_a^\ast +  \sum_{bcd}^{\in\{f,g\}}\kappa_{gbcd}\chi_b^\ast\chi_c^\ast\chi_d^\ast,\label{eqn:g_eom_2}
\end{multline}
which only contains the modes that we numerically integrate.
The second term is $\Delta_\beta \chi_g\equiv -E_0^{-1}\partial (H^{\beta}_{\rm int} + H^{\beta}_{\rm ext})/\partial \chi_g^\ast$,
\begin{align}
    &\Delta_\beta \chi_{g}= 2(\kappa_{gg\beta_{-4}}\chi_g^\ast \chi_{\beta_{-4}}^\ast +\kappa_{gg_0\beta_{-2}}\chi_{g_0}^\ast\chi_{\beta_{-2}}^\ast\nonumber
\\
&\phantom{\Delta_\beta \chi_{g}= 2(}+\kappa_{gg^\ast\beta_0}\chi_g\chi_{\beta_0}^\ast
+\kappa_{gf\beta_{-4}}\chi_f^\ast \chi_{\beta_{-4}}^\ast\nonumber \\
&\phantom{\Delta_\beta \chi_{g}= 2(}+\kappa_{gf_0 \beta_{-2}}\chi_{f_0}^\ast\chi_{\beta_{-2}}^\ast
+\kappa_{gf^\ast \beta_0}\chi_f\chi_{\beta_0}^\ast)\nonumber\\
&\phantom{\Delta_\beta \chi_{g}=}+U_{g\beta_{-4}}^\ast\chi_{\beta_{-4}}^\ast+U_{g\beta_{-2}}^\ast\chi_{\beta_{-2}}^\ast+U_{g\beta_{0}}^\ast\chi_{\beta_{0}}^\ast.\label{eqn:p2}
\end{align}
At each time step, we compute $\Delta_\beta \chi_g$ using Eqs.~(\ref{eqn:p2}) and (\ref{eqn:p5})--(\ref{eqn:p7}). 
Afterward, we use it to drive the $g$-mode according to Eqs.~(\ref{eqn:g_eom_1}) and (\ref{eqn:g_eom_2}). 

\subsection{Numerical Result\label{subsec:num}}
In this section, we describe how to numerically evolve the inspiral accounting for the excitation of $l=2$ $f$- and $g$-modes, as well as the adiabatically sourced modes. 
In the previous section, we derived evolution equations for the mode amplitudes. 
To close the system, we need evolution equations for the orbital elements. 
In order to do so, we need to model the backreaction of the modes on the orbit, which we can achieve using the Hamiltonians outlined in the previous section. 

The evolution equations for the orbital elements $D$ (semi-major axis) and $\Phi$ (orbital phase) take the form:
\begin{align}
    &\ddot{D}-D\dot{\Phi}^2+\frac{GM_{\rm t}}{c^2D}=g_D\label{eqn:eom_D}\\
    &D\ddot{\Phi}+2\dot{D}\dot{\Phi}=g_\Phi,\label{eqn:eom_phi}
\end{align}
where $M_{\rm t}$ is the total mass of the binary, and the terms $g_D$ and $g_\Phi$ represent the radial and azimuthal accelerations. The accelerations have contributions due to gravitational wave emission and tidal excitation, \emph{i.e.}, for $i\in \{D,\Phi\}$
\begin{align}
    &g_i = g_i^{\rm (gw)}+ g_i^{\rm (tide)}, \label{eq:gis}
\end{align}
where we use Eqs.~(28), (29), (31), and (35) from Yu~\emph{et~al.}~\cite{yu23} for $g_i$ while accounting for more modes than those included therein.
In our numerical calculations, we assume that the companion NS is identical to the primary, and account for tidal effects in both NSs. 
When computing $g_{i}^{\rm (gw)}$, we ignore terms from the tidally induced NS quadrupoles (we ignore Eqs.~C26 and C28 in Yu~\emph{et~al.}~\cite{yu23}). 

\begin{figure}
    \centering
    \includegraphics[width=0.48\textwidth]{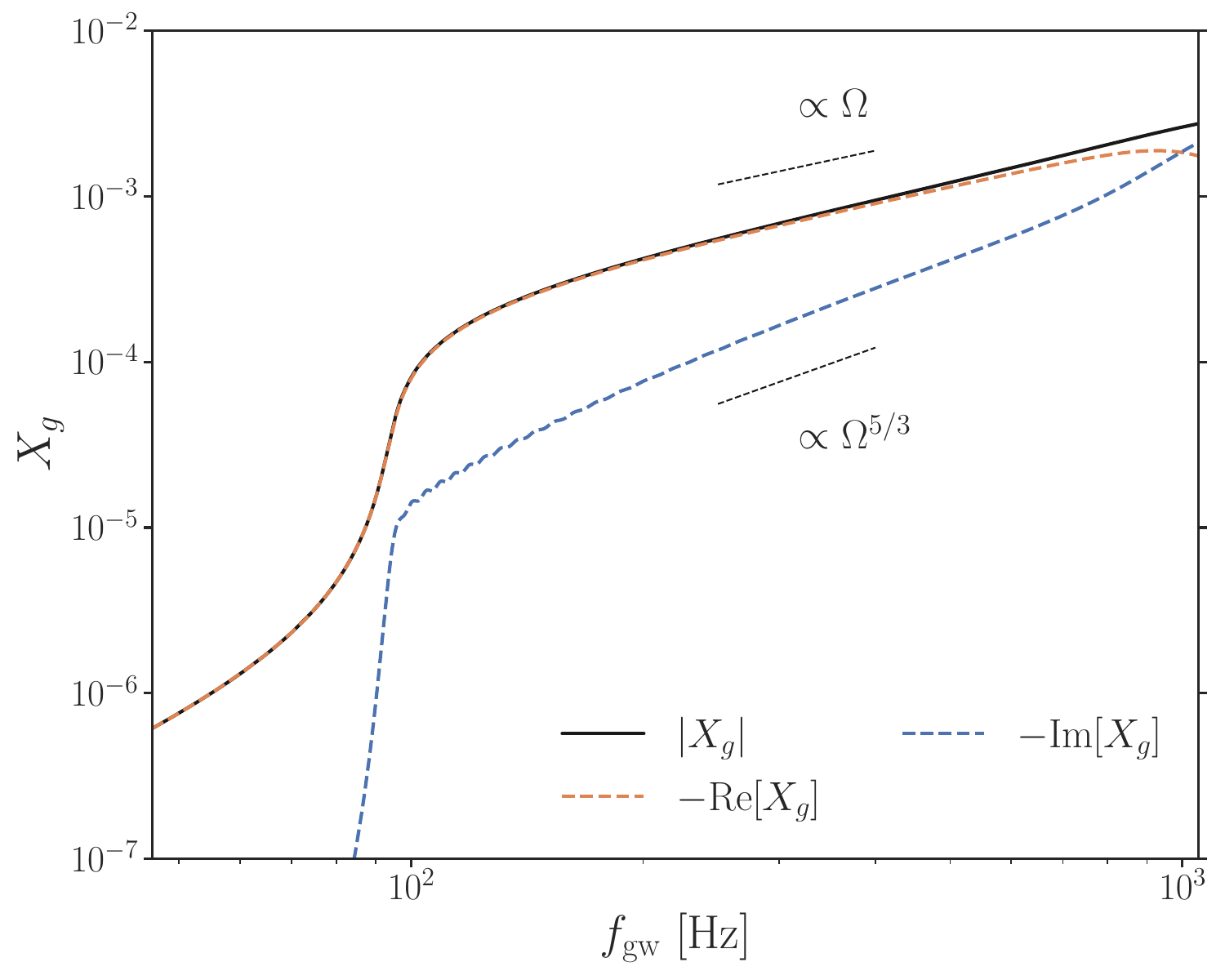}
    \caption{Evolution of the nonlinearly driven amplitude of $n=1$, $l=m=2$ $g$-mode. We show its absolute magnitude~(black), along with its real~(orange) and imaginary parts~(blue). As shown by \citetalias{nlrl_prl}, the real and imaginary components of $X_g$ are locked to $\Omega$ and $\Omega^{5/3}$ approximately. At $f_{\rm gw}\approx 1\,{\rm kHz}$, the imaginary part becomes comparable to the real part and eventually exceeds it. 
    Thus, the locking condition in Eq.~(\ref{eqn:break_cond}) no longer holds and the lock starts to break.}
    \label{fig:full_sol}
\end{figure}
We present the amplitude of the nonlinearly driven $n=1$, $l=m=2$ $g$-mode throughout the inspiral in Fig.~\ref{fig:full_sol}. We show the absolute magnitude of the amplitude~(black), along with its real~(orange) and imaginary~(blue) parts. 

As we discussed in Sec.~\ref{subsec:simp}, the real and imaginary parts of $X_g$ are approximately proportional to $\Omega$ and $\Omega^{5/3}$. At $f_{\rm gw}\approx 1\,{\rm kHz}$, the lock condition in Eq.~(\ref{eqn:break_cond}) no longer holds. 
However, the lock breaks slowly due to the presence of an imaginary component of the frequency shift. 
As a result, the resonantly locked $g$-mode grows monotonically and coherently with the orbit for most of the inspiral.

\begin{figure}
    \centering
    \includegraphics[width=0.48\textwidth]{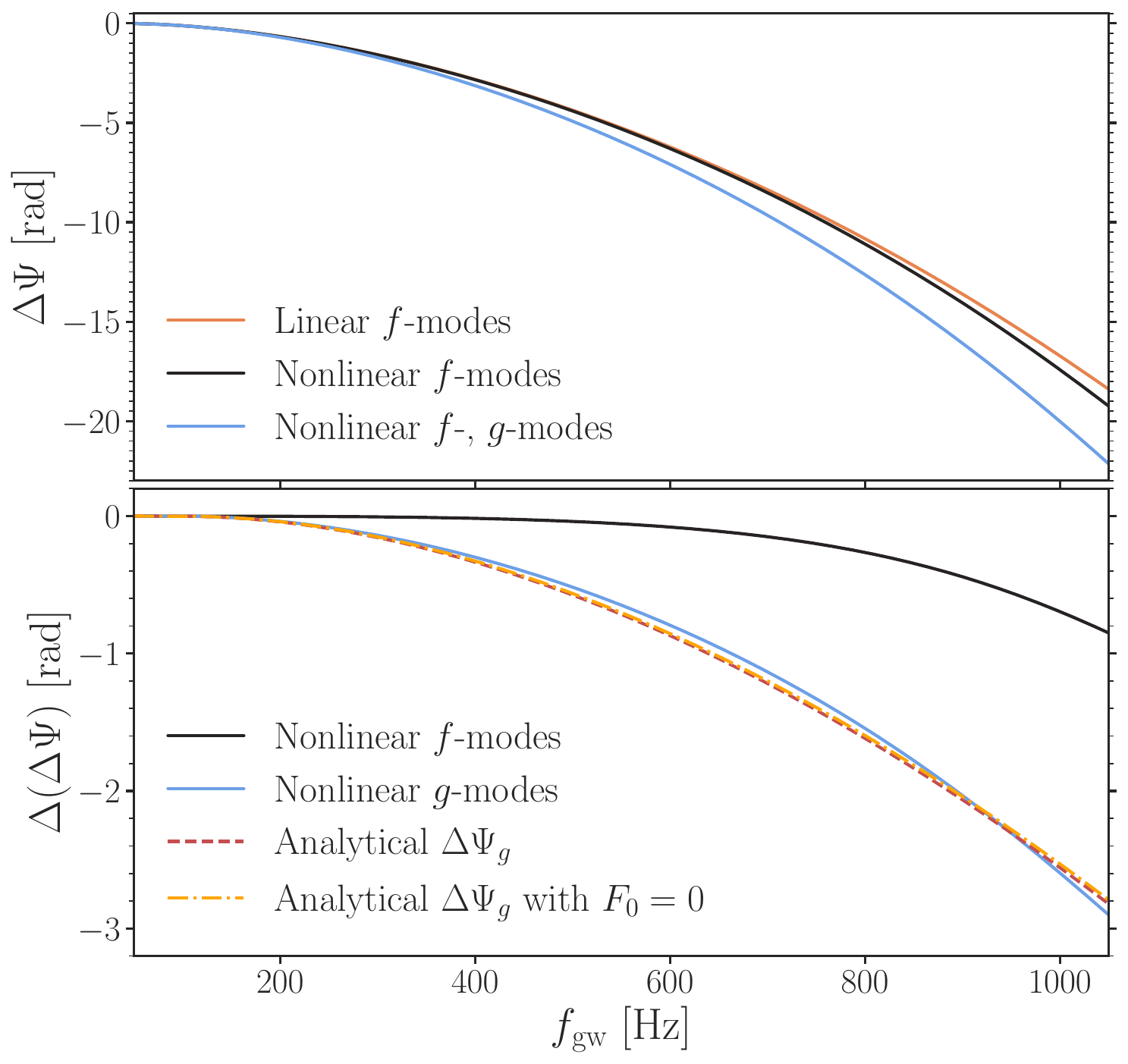}
    \caption{Tidal dephasing on the frequency-domain GW computed by numerically integrating the inspiral using Eqs.~(\ref{eqn:eom_D}) and (\ref{eqn:eom_phi}). 
    We include $n=1$, $l=2$ $g$-modes, as well as $l=0$, $l=2$, and $l=4$ $f$- and $p$-modes up to radial order $n=60$.
    \emph{Top}---Dephasing due to linear $f$-modes (orange), nonlinear $f$-modes (black), and nonlinear $f$- and $g$-modes (blue).
    \emph{Bottom}---Dephasing solely due to nonlinear $f$-modes (black), nonlinear $g$-modes (blue).
    We show our analytical estimation of dephasing due to the resonantly locked $g$-modes, given in Eq.~(\ref{eqn:deph_psi}) (red).
    We additionally show $\Delta \Psi_g$ ignoring terms that contain $F_0$ (orange).
    The terms with $F_0$ have practically negligible impact on $\Delta\Psi_g$.
    }
    \label{fig:dephasing}
\end{figure}

In Fig.~\ref{fig:dephasing}, we demonstrate the impact of RL on the GW signal's phase in the frequency domain. 
In the top panel, we show the baseline system with only a linearly-driven $f$-mode (orange), the cumulative effects of accounting for nonlinearity in $f$-modes (black), and additionally including nonlinear $g$-modes. 
The dephasing is computed with respect to the point particle inspiral $\Delta \Psi\equiv \Psi - \Psi_{\rm pp}$.
In the bottom panel, we isolate each effect by subtracting the $\Delta \Psi$ of the linear $f$-mode system from that of nonlinear $f$-mode (black). Similarly, we subtract $\Delta \Psi$ of nonlinear $f$-mode from that of nonlinear $f$- and $g$-modes (blue). 
This illustrates that the resonantly locked $g$-modes lead to an additional phase correction of $\mathcal{O}(3)\,{\rm rad}$ near the merger.
This phase correction is significantly larger than what is expected from the linear theory, in which the resonant $g$-modes interact with the orbit only briefly during a narrow frequency range. 
Therefore, neglecting the nonlinearity in $g$-modes can potentially bias the inference of physical properties of NSs from GW signals.

In the following section, we will use an energy balancing argument to analytically estimate the dephasing due to RL on the frequency-domain GW. 
We show the resulting analytical estimate~($\Delta \Psi_g$) in the bottom panel (red dashed), which is in good agreement with the numerical result. 
Our full expression for $\Delta\Psi_g$ in Eq.~\eqref{eqn:f0} includes terms arising from the initial energies of $g$-modes at the onset of RL (when $2\pi f_{\rm gw}=\omega_g$). 
In the orange curve, we show the result of neglecting the phasing effect of the $g$-modes' initial energies in the expression of $\Delta\Psi_g$.
The comparison between the red and orange curves shows that we can neglect the effect of initial energies for practical purposes, which can simplify the analytical description.
The phase shift comes with an associated time shift; we refer readers to Fig.~4 of \citetalias{nlrl_prl} [this time shift is already included in the estimate of the frequency domain phase shift via Eq.~\eqref{eqn:deph3}].

\subsection{Analytical Dephasing Estimate\label{subsec:deph_est}}
We have shown that the $g$-modes grow monotonically while staying in a near-resonant state and that they cause significant dephasing of the GW signal. 
In this section, we will come up with an analytical estimate of the dephasing due to the resonantly locked $g$-modes, which can help with including this effect in data analysis.

Consider two binaries where only one~(labeled with subscript 2) has resonantly locked $m\pm 2$ $g$-modes in a single NS, while in the other binary, the masses are treated as point particles. 
Assume the binaries are otherwise identical. 
RL leads to a time shift at a given $f_{\rm gw}$ (\emph{i.e.}, a shift in the time at which the GW signal reaches a particular frequency $f_{\rm gw}$), represented by $\Delta t(f_{\rm gw})=t_2(f_{\rm gw})-t_1(f_{\rm gw})$. 
At these times, the phases of the time-domain GW of the two systems are $\Phi_{{\rm gw,1}}$ and $\Phi_{{\rm gw},2}=\Phi_{{\rm gw,1}} + \Delta \Phi_{\rm gw}$ (this is a slight abuse of notation because the time-domain phases on the left- and right-hand-side are evaluated at different times, \emph{i.e.}, $\Phi_{{\rm gw},2}\left[ t_2(f_{\rm gw}) \right] = \Phi_{{\rm gw,1}} \left[ t_1(f_{\rm gw}) \right] + \Delta \Phi_{\rm gw}$. This is conventionally left implicit and the phase difference is written as a function of the GW frequency $\Delta \Phi_{\rm gw}(f_{\rm gw})$ to indicate this).

Letting the energy of the two systems as $E_1$ and $E_2=E_1 + \Delta E$, we can calculate $\Delta \Phi_{\rm gw}$ and $\Delta t$ using~\cite{yu23} 
\begin{align}
   & \frac{{\rm d}\Delta \Phi_{\rm gw}}{{\rm d}f_{\rm gw}}\approx 2\pi f_{\rm gw}\frac{1}{\dot{E}_1}\frac{{\rm d}\Delta E}{{\rm d} f_{\rm gw}},\label{eqn:deph1}\\
   &\frac{{\rm d}\Delta t}{{\rm d}f_{\rm gw}}=\frac{1}{\dot{E}_1}\frac{{\rm d}\Delta E}{{\rm d}f_{\rm gw}},\label{eqn:deph2}
\end{align}
where we have ignored the enhancement in radiation in Eq.~(\ref{eqn:deph1}) since the resonantly locked $g$-modes primarily modify the orbital dynamics through a conservative energy transfer [see Eq.~(84) of Yu \emph{et~al.}~\cite{yu24}]. 
Finally, in the stationary-phase approximation, the dephasing of the frequency domain waveform is given by \cite{1994PhRvD..49.2658C}
\begin{align}
    \Delta \Psi(f_{\rm gw}) = 2\pi f_{\rm gw}\Delta t (f_{\rm gw})-\Delta \Phi_{\rm gw} (f_{\rm gw}),\label{eqn:deph3}
\end{align}
According to Eq.~\eqref{eqn:deph1}, the time-domain phase shift needs the change in the system's energy $\Delta E$; the change in the energy due to the locked $g$-modes in a single NS is (see Appendix~\ref{app:nl}) 
\begin{align}
    &\frac{\Delta E}{E_0} \approx \left(\frac{|\dot{\chi}_g|^2}{\omega_g^2} + |\chi_g|^2 -\frac{1}{2}\kappa_{\rm eff}|\chi_g|^4\right).\label{eqn:deph4}
\end{align}
In principle, $\Delta E$ also contains the interaction energy between the modes and the orbit and the tidal modification to the orbital energy. 
However, since the $g$-modes stay on resonance, they mainly modify the tidal torque rather than the orbital energy due to the radial interaction (\emph{e.g.}, see Fig.~3 of Yu~\emph{et~al.} \cite{yu24}) and hence we can restrict to the mode energy in Eq.~\eqref{eqn:deph4}.

We can evaluate the second and the third terms on the right-hand side of Eq.~(\ref{eqn:deph4}) using Eq.~(\ref{eqn:locked_amp}), 
\begin{align}
    |X_g|^2=|\chi_g|^2 \approx \frac{\omega_g^2 - 4\pi^2f_{\rm gw}^2}{\omega_g^2\kappa_{\rm eff}} + \bigg(\frac{V_g|_{2\Omega=\omega_g}}{\kappa_{\rm eff}}\bigg)^{2/3}
\end{align}
for $2\pi f_{\rm gw} \geq \omega_g$. 
The kinetic energy can be estimated as $|\dot{\chi}_g|^2  \approx 4\Omega^2|X_g|^2$. Plugging these expressions in Eqs.~(\ref{eqn:deph1})--(\ref{eqn:deph4}) and using the initial values
\begin{align}
&\Delta \Phi(f_{\rm gw}=\omega_g/2\pi)=\frac{2\pi f_{\rm gw}\Delta E}{\dot{E}_1}\bigg|_{f_{\rm gw}=\omega_g/2\pi},\label{eqn:init_cond1}\\
&\Delta t(f_{\rm gw}=\omega_g/2\pi)= \frac{\Delta E}{\dot{E}_1}\bigg|_{f_{\rm gw}=\omega_g/2\pi},\label{eqn:init_cond2}
\end{align}
we obtain the expression for $\Delta \Psi$, 
\begin{align}
 &\Delta \Psi_g \approx K[54a^5-5(21 + 10 F_0 -F_0^2)a^3+96+140 F_0\nonumber \\
 &\phantom{\Delta\Psi_g\approx K}- 5 F_0^2-45(1  +2F_0 )a^{-1}]\Theta(a-1),\label{eqn:deph_psi}\\
 &F_0 =-\bigg|\frac{3\pi I_g^2\kappa_{\rm eff}}{10}\bigg|^{1/3}\bigg(\frac{\omega^2_gR^3}{4GM_{\rm t}}\bigg)^{2/3},\label{eqn:f0}\\
&a = \left(\frac{2\pi f_{\rm gw}}{\omega_g}\right)^{1/3}=
\bigg(\frac{GM_{\rm t}\omega_g}{2c^3}\bigg)^{-1/3}v\label{eqn:deph_a},\\
&K=\frac{1}{2^{8/3}\kappa_{\rm eff}\eta^2}\frac{E_0}{M_{\rm t} c^2}\bigg(\frac{G M_{\rm t}\omega_g}{c^3}\bigg)^{-7/3},\label{eqn:deph_K}
\end{align}
where $v\equiv (\pi f_{\rm gw} M_{\rm t}G/c^3)^{1/3}$ and  $\eta=MM'/M_{\rm t}^2$ is the symmetric mass ratio. 
We have taken $\dot{E}_1=-(32\pi/5)\eta^2(\pi GM_{\rm t}f_{\rm gw}/c^3)^{7/3}(M_{\rm t} c^2)f_{\rm gw}$. 
The terms that contain $F_0$ in Eq.~(\ref{eqn:deph_psi}) arise from the initial 
energy of $g$-modes at $2\pi f_{\rm gw}=\omega_g$ in Eqs.~(\ref{eqn:init_cond1}) and (\ref{eqn:init_cond2}). 
Their phasing effect is distinguishable from the linear resonance due to the term proportional to $a^{-1}$.
The result of the numerical calculation in Fig.~\ref{fig:dephasing} shows that their contribution is small and may practically be neglected. 
Equation~(\ref{eqn:deph_psi}) is a contribution from a single NS, the effect of RL in the companion star can be obtained by appropriately replacing $E_0$, $\omega_g$, $I_g$, and $\kappa_{\rm eff}$ with the companion's values.

Eqs.~(\ref{eqn:f0}) and (\ref{eqn:deph_K}) imply that the dephasing due to RL is a function of $\omega_g$, $I_g$, and $\kappa_{\rm eff}$. 
In particular, $\omega_g$ depends on the Brunt-V\"ais\"al\"a frequency of NS~\cite{reisenneger92, aerts10}. Therefore, if the dephasing due to RL is measurable, it can deliver information about the stratification of NSs.

\section{Discussion\label{sec:disc}}

In the crust of our NS where the Brunt-V\"ais\"al\"a frequency vanishes, the $g$-modes rapidly become evanescent. Thus, the growth of the $g$-mode induces the shear motion at the crust-core boundary, which could act as a source of damping due to the viscous boundary layer~(VBL) damping. 
Here, we discuss whether the viscous damping can break the lock. 
The spirit of the analysis is similar to that of Bildsten and Ushomirsky~\cite{bildsten00}, where the oscillatory motion of $r$-modes at the VBL provided the damping source.
We simply account for the fact that the motion is now monotonic.

To estimate the energy transfer between the mode and VBL, we assume that the kinematic viscosity $\nu$ takes a value of $1.8\times 10^4\,{\rm cm}^2\,{\rm s}^{-1}/T_8^2$ where $T_8=T/10^{8}\,{\rm K}$. Then, the thickness of VBL, $\delta$, is given by
\begin{align}
    \delta\approx \sqrt{\tau \nu},
\end{align}
where $\tau$ is the growth timescale of the mode, $X_g/\dot{X}_g$. Given $\delta$, we can approximately obtain what fraction of the mode energy is contained within VBL, and thus damped. 
At $T=10^6\,{\rm K}$, we estimate $75$\% of the $g$-mode energy is damped.

Our result presented in Fig.~\ref{fig:full_sol} indicates that the two resonantly locked $g$-modes~($m=\pm 2$) has the energy of $0.75\times 2E_0|X_g|^2\approx 10^{-4}E_0\sim 3\times 10^{49}\,{\rm ergs}$ at the contact. 
The mass of the crust in our NS model is approximately $10^{32}\, {\rm g}$ and the energy deposited per crustal mass is therefore $3\times 10^{17}\,{\rm ergs}\,{\rm g}^{-1}=k_BT/m_n$, where $m_n$ is the rest mass of a neutron. 
Therefore, the VBL is heated to $T\sim 10^{10}\,{\rm K}$, which in turn will significantly suppress the viscous damping due to the lower values of $\nu$. Hence, we conclude that the shear dissipation in VBL is a self-limiting effect, which is unlikely to affect our analysis.

RL is a consequence of nonlinear hydrodynamic interactions within NSs, which cause $g$-modes to oscillate at frequencies much higher than their natural frequencies.
Another nonlinear phenomenon that may arise during NS mergers is the tidal $p$-$g$ instability~\cite{w13, vzh, w16}.
Below, we distinguish the two effects.

In contrast to RL, where the dynamics are primarily driven by the $gggg$ (and $gggf$) couplings, the $p$-$g$ instability involves interactions between two short-wavelength daughter modes and two long-wavelength parent modes, typically via the $ggff$ channel. This coupling is intrinsically different 
than the $gggg$ coupling, as $\kappa_{gg^\ast ff^\ast} \sim [(1/r) {\rm d} (\xi_{g^\ast}^{\rm h} \xi_f^{\rm h})/{\rm  d}r]^2 \sim \omega_g^{-4}$, while $\kappa_{gg^\ast gg^\ast} \sim \omega_g^{-6}$.
Its effect is to lower $\omega_{\rm eff}^2$ to a negative value, which results in the exponential growth of the daughter mode (whereas RL increases $\omega_{\rm eff}^2$). The $p$-$g$ instability is more likely to happen in high-order (low-frequency) $g$-modes, whereas the RL happens only for the lowest order $g$-mode.
Therefore, despite both phenomena stemming from nonlinear interactions, they are fundamentally different and have distinct effects on GW signatures.

\section{Conclusion\label{sec:conc}}

We have presented the details behind the RL of $g$-modes in coalescing NS binaries, originally demonstrated by \citetalias{nlrl_prl}. Using a $1.4$-$1.4\,{\rm M_\odot}$ Newtonian NS binary constructed using SLy4 EOS, we have computed the evolution of $n=1$, $l=m=2$ $g$-modes. We have incorporated nonlinear wave interaction up to quartic order and the quadratic tidal driving with $l=0,2,4$ $f$- and $p$-modes up to $n=60$. 

We have demonstrated that the lowest-order $g$-modes receive significant nonlinear frequency shift, which leads to the anharmonicity; \emph{i.e.}, their oscillation frequency depends on the energy of the mode. RL occurs as a direct consequence of this anharmonicity; as the tidal forcing frequency approaches the $g$-mode's linear resonance point ($2\Omega=\omega_g$), the mode's amplitude, thus its energy, grows. Due to the anharmonicity, the $g$-mode frequency shifts upward along with the tidal forcing frequency, creating a self-sustaining lock (or a feedback loop) that keeps the $g$-mode in a resonantly amplified state. 
We stress that RL represents a component of the truly dynamical tide since its phenomenology depends on the history of the mode excitation.

To describe this effect, we have parameterized the nonlinearly corrected frequency using $\omega_{\rm eff}(\chi_g;\kappa_{\rm eff}, \gamma_{\rm eff})$.
The effective coupling coefficient $\kappa_{\rm eff}$ parameterizes the effective four-mode coupling between $l=|m|=2$ $g$-modes with radial order $n=1$, which endows the $g$-mode with the anharmonicity. 
The initiation of RL depends on whether $\kappa_{\rm eff}$ is greater than the critical value that we show in Eq.~(\ref{eqn:kap_crit}).
This condition is demonstrated in Fig.~\ref{fig:kap_crit}.
The effective (anti-) damping coefficient $\gamma_{\rm eff}$ is the consequence of energy transfer between $f$- and $g$-modes.
This interaction results in the imaginary component of the nonlinearly corrected frequency, which can modify the damping of the mode amplitude induced by the inspiral. 
The lock can continue until the end of the merger if $\gamma_{\rm eff}I_fI_g>0$ and $|\gamma_{\rm eff}|$ exceeds the critical value $\gamma_{\rm crit}$ in Eq.~(\ref{eqn:gam_crit}). 
We have illustrated this condition in Fig.~\ref{fig:gam_crit}.
 
Our numerical result in Fig.~\ref{fig:full_sol} agrees with the analytical expectation laid out by \citetalias{nlrl_prl} and explained in detail in this paper, where both the real and the imaginary parts of the amplitude evolve proportionally to $\Omega$ and $\Omega^{5/3}$. 
This implies the universality of RL across different choices of the nuclear equation of state on which the exact values $\kappa_{\rm eff}$ and $\gamma_{\rm eff}$ depend. 
We have also shown in Figs.~\ref{fig:kap_crit} and \ref{fig:kap_crit2} and numerically verified that RL is only significant for $n=1$, $l=|m|=2$ $g$-modes.

As the resonantly locked $g$-modes oscillate phase-coherently with the orbit, it impacts the observable GW signal. Due to its monotonic growth, the locked $g$-modes can grow to significant energy.
Our numerical calculation yields that the dephasing on the frequency-domain GW accumulates to $\mathcal{O}(3)\,{\rm rad}$ at $f_{\rm gw}=1.05\,{\rm kHz}$, as shown in Fig.~\ref{fig:dephasing} (and Fig.~4 of \citetalias{nlrl_prl}).

As the significance of $g$-modes has thus far been overlooked due to its subdominant tidal coupling, many NR simulations have utilized barotropic equations of state that lack compositional dependence. This removes $g$-modes from the simulations as they are driven by the buoyancy due to compositional gradient~\cite{reisenneger92}. Our analysis indicates that neglecting $g$-modes can lead to inaccurate GW phases calculated from NR simulations. As many waveform approximants are calibrated against NR simulations, this can bias the inference of NS bulk properties from observed GW signals, such as tidal deformability.

Furthermore, inspection of Eqs.~(\ref{eqn:deph_psi})--(\ref{eqn:deph_K}) implies that if $\Delta\Psi_g$ is measurable, we can potentially obtain information on the stratification within NSs. Note that both the composition and sound speeds that affect the stratification depend on the nuclear EOS. 
Therefore, the signature of RL can potentially inform a model EOS if measured.

We note that our analysis is done in Newtonian order. To account for general relativistic effects, one would solve for the NS structure using the Tolman-Oppenheimer-Volkoff equation and for quasi-normal modes using the relativistic Euler equations~(see, \emph{e.g.}, Ref.~\cite{ferrari2008} for review). This will modify the compactness of the star as well as the $g$-mode eigenfrequency measured in the inertial frame. In addition, to account for nonlinearity, we require relativistic coupling coefficients, most importantly $\kappa_{\rm eff}$ that RL depends on. 

Also, we have assumed normal fluid NSs composed of neutrons, protons, and electrons. In reality, evidence suggests that the inner core of the NSs is in a superfluid state. Yu and Weinberg~\cite{yu17} have demonstrated that, given the same radial orders, $g$-modes in superfluid NSs have a denser spectrum and larger frequencies. Therefore, the superfluidity may alter the impact of RL on GW phasing which depends on the $g$-mode eigenfrequencies.

As the $g$-modes are excited to higher amplitudes, significant energies can be deposited in the crust. Our preliminary work in Sec.~\ref{sec:disc} indicates that the crust-core boundary can be heated to a significant temperature $T\sim 10^{10}\,{\rm K}$, which can result in partial crust melting. Furthermore, depending on the exact composition of the crust, the deposited energy may lead to richer phenomena such as crustal shattering or EM signatures, which can significantly modify the $g$-mode dynamics.

\section*{Acknowledgment}
The authors appreciate Lars Bildsten for assisting with the work on VBL in Sec.~\ref{sec:disc}. The authors have also benefited from discussions with Eliot Quataert and Christopher Hirata.
HY acknowledges support from NSF grant No. PHY-2308415 and from Montana NASA EPSCoR Research Infrastructure Development under award No. 80NSSC22M0042. 
TV acknowledges support from NSF grants 2012086 and 2309360, the Alfred P. Sloan Foundation through grant number FG-2023-20470, the BSF through award number 2022136, and the Hellman Family Faculty Fellowship.

\appendix
\section{Nonlinear Terms in $L_{\ast}$\label{app:nonlin}}
Consider an NS of mass $M$ and $R$ in hydrostatic equilibrium.
The Lagrangian that describes the intrinsic energy of the star is
\begin{align}
    L_{\rm int }&=\int {\rm d}^3x\rho(\mbf{x})\,\bigg[\frac{1}{2}|\dot{\mbf{x}}|^2 - u(\tau,\bm{\mu}) - \frac{1}{2}\Phi_\ast(\mbf{x})\bigg], \label{eqn:l_int_1}
\end{align}
where $\rho$ is the background density, $\dot{\mbf{x}}\equiv{\rm d}\mbf{x}/{\rm d}t$, $u$ is the specific internal energy, $\tau$ is the specific volume, $\bm{\mu}$ is a compositional parameter, and $\Phi_\ast$ is the gravitational potential
\begin{align}
    \Phi_\ast(\mbf{x})=-G\int {\rm d^3}y\,\frac{\rho(\mbf{y})}{|\mbf{x}-\mbf{y}|}.
\end{align}

When the NS is in binary, the time-dependent tidal potential generated by the companion displaces a fluid element in the primary originally at $\mbf{x}$ to $\mbf{x}'=\mbf{x}+\bm{\chi}(t,\mbf{x})$.
Simultaneously, $U$ perturbs $\rho$, $u$, and $\Phi_\ast$ as well.
We can write the Lagrangian of the perturbed NS~($L_\ast'$) in terms of the intrinsic and external components,
\begin{align}
    L_\ast' = L_{\rm int}' + L_{\rm ext}',
\end{align}
where the prime symbol suggests that the star is perturbed.

Assuming the NS is originally at static equilibrium, the intrinsic part $L_{\rm int}'$ reads
\begin{align}
    L_{\rm int}'&=\int {\rm d}^3x'\rho'(\mbf{x}')\,\bigg[\frac{1}{2}|\dot{\mbf{x}}'|^2 - u'(\tau', \bm{\mu}') - \frac{1}{2}\Phi_\ast'(\mbf{x}')\bigg] \nonumber\\
    &=\int {\rm d}^3x\rho(\mbf{x})\,\bigg[\frac{1}{2}|\dot{\bm{\chi}}|^2 - u'(\tau', \bm{\mu}') - \frac{1}{2}\Phi_\ast'(\mbf{x}')\bigg].\label{eqn:l_int_3}
\end{align}
To obtain the second equality, we have utilized the fact that the mass of the fluid element is conserved, \emph{i.e.}, ${\rm d}^3x'\,\rho'(\mbf{x}')={\rm d}^3x\, \rho(\mbf{x})$.
We assume that the oscillation is adiabatic by letting $\bm{\mu}'=\bm{\mu}$.

Using the definition of Lagrangian perturbation ($\Delta$) of a given quantity $f$,
\begin{align}
    f'(t,\mbf{x}')=f(t,\mbf{x})+\Delta f(t,\mbf{x}'),
\end{align}
we can rewrite Eq.~(\ref{eqn:l_int_3}) as
\begin{align}
    L_{\rm int}'
    &=\int {\rm d}^{3}x\rho(\mbf{x})\,\bigg(\frac{1}{2}|\dot{\bm{\chi}}|^2 - \Delta u - \frac{1}{2}\Delta\Phi_\ast\bigg) + L_{\rm int}.
\end{align}

Furthermore, 
$\Delta u$ and $\Delta \Phi_\ast $ can be expanded as
\begin{align}
    &\Delta u (\tau', \bm{\mu})=\sum_{n=0}\bigg(\frac{\partial^n u'}{\partial \tau^n}\bigg)_{\bm{\mu}}(\delta \tau)^n-u(\tau, \bm{\mu})\nonumber \\
    &\phantom{\Delta u (\tau, \bm{\mu})}=\delta u(\tau, \bm{\mu}) + \sum_{n=1}\bigg(\frac{\partial^n u'}{\partial \tau^n}\bigg)_{\bm{\mu}}(\delta \tau)^n,\\
    &\Delta \Phi_\ast(\mbf{x}') = \delta\Phi_\ast(\mbf{x}) + \sum_{N=1}\frac{\chi^N\nabla_N\Phi_\ast'}{N!},
\end{align}
where we have used the notation of Poisson and Will~\cite{poisson14} to write $\chi^N\nabla_N=\chi^{i_1}\chi^{i_2}\dotsi\chi^{i_N}\nabla_{i_1}\nabla_{i_2}\dotsi\nabla_{i_N}$.
Here, the terms in the summation encode the internal, conservative, and nonlinear interaction of the Lagrangian displacement $\bm{\chi}$. 
Taking the action $\int {\rm d}t\, L_{\rm int}'$ and minimizing it with respect to $\delta \bm{\chi}$ yields the nonlinear internal acceleration terms $\mbf{a}_{\rm int}^{(i)}$ in Eq.~(\ref{eqn:eom_chi}).

Meanwhile, the expression for $L_{\rm ext}'$ is
\begin{align}
    L_{\rm ext}'&=-\int {\rm d}^3x'\,\rho'(\mbf{x}')U(\mbf{x}'),\label{eqn:lag_ext}
\end{align}
where $U$ is the tidal potential generated by the companion of mass $M'$ 
\begin{align}
    U(\mbf{x})&=-\frac{GM'}{|\mbf{x}-\mbf{D}|}.
\end{align}
Here, $\mbf{D}$ is the orbital separation vector connecting the centers of masses of the binary components. 
In practice, we additionally expand $U$ in terms of spherical harmonics to obtain,
\begin{align}
U=-GM'\sum_{lm}W_{lm}\bigg(\frac{R}{D}\bigg)^{l+1} \frac{r^l}{R^{l+1}}  e^{-im\Phi} Y_{lm}(\theta,\phi),\label{eqn:tidal_u}
\end{align}
where $r=|\mbf{x}|$, $D=|\mbf{D}|$, and $W_{lm}=4\pi(2l+1)^{-1} Y_{lm}(\pi/2,0)$.
This form is useful in extracting the $l=2$ harmonics that dominate the tidal interaction.

We expand the external Lagrangian as
\begin{align}
    L_{\rm ext}'\approx -\int {\rm d}^3x\,\rho(\mbf{x})\Bigg[U(\mbf{x})+\sum_{N=1}\frac{\chi^N\nabla_N U}{N!}\bigg|_{\mbf{x}'=\mbf{x}}\Bigg],
\end{align}
and minimize the action $\int {\rm d}t\, L_{\rm ext}'$ with respect to $\delta \bm{\chi}$ to obtain the external acceleration terms $\mbf{a}_{\rm ext}^{(i)}$.

\section{Eigenmodes of Stellar Oscillation\label{app:eigen}}
We have described tidal deformation on NS using eigenmodes.
In this section, we briefly explain how we obtain the eigenfrequency and spatial eigenfunction that characterize each mode (for extensive details, see, \emph{e.g.}, Unno \emph{et~al.}~\cite{unno89}).

Consider a steady flow with density $\rho(\mbf{x})$ and velocity $\mbf{v}\equiv {\rm d}\mbf{x}/{\rm d}t$.
For an arbitrary closed volume $V$ with the surface $S$, the rate of change in the mass in $V$ is equal to the mass flow through $S$. 
This gives the equation of mass continuity,
\begin{align}
    \frac{\partial \rho}{\partial t} + \nabla\cdot (\rho\mbf{v})=0.\label{eqn:mass_cons}
\end{align}
Furthermore, varying the action $S_{\rm int}=\int {\rm d}t\,L_{\rm int}$ with respect to $\mbf{x}$ yields
\begin{align}
    \frac{\delta S_{\rm int}}{\delta \mbf{x}}&=\iint{\rm d}t\, {\rm d}^3x \rho(\mbf{x})\,\bigg[-\ddot{\mbf{x}}-\left(\frac{\partial u}{\partial \tau}\right)_s\frac{\partial \tau}{\partial \mbf{x}}-\frac{\partial\Phi_\ast}{\partial \mbf{x}}\bigg]\nonumber\\
    &=\iint{\rm d}t\, {\rm d}^3x \rho(\mbf{x})\bigg(-\ddot{\mbf{x}}-\frac{\partial P}{\partial \mbf{x}}\rho^{-1}-\frac{\partial\Phi_\ast}{\partial \mbf{x}}\bigg),
\end{align}
where $\ddot{\mbf{x}}={\rm d}^2\mbf{x}/{\rm d}t^2$, $s$ is the specific entropy, and $\tau=\rho^{-1}$ is the specific volume.
To obtain the second equality, we have used integration by parts and used the fact the pressure vanishes in the boundary.
Letting $\delta S_{\rm int}/\delta \mbf{x}=0$ leads to the conservation of momentum,
\begin{align}
    \rho\ddot{\mbf{x}}=-\nabla P-\rho\nabla\Phi_\ast.\label{eqn:momentum_cons}
\end{align}
Note that $P$ and $\rho$ are related by an equation of state
\begin{align}
    P=P(\rho, \bm{\mu}),\label{eqn:eos}
\end{align}
where $\bm{\mu}$ is a compositional parameter, which in our case is the proton fraction for $\beta$-equilibrium.
The fluid element also satisfies the Poisson's equation
\begin{align}
    \nabla^2\Phi_\ast = 4\pi G\rho.\label{eqn:poisson}
\end{align}
Eqs.~(\ref{eqn:mass_cons}), (\ref{eqn:momentum_cons}), (\ref{eqn:eos}), and (\ref{eqn:poisson}) form a closed system of equations that governs the motion of the fluid element.

Now, we consider a small, adiabatic perturbation on a spherically symmetric, non-rotating star in a hydrostatic equilibrium. 
It displaces a fluid element in the star at $\mbf{x}$ and causes perturbations on $\rho$, $P$, and $\Phi_\ast$ around their equilibrium states.
\begin{align}
    &\rho'(t,\mbf{x})=\rho(r)+\delta\rho(t,\mbf{x}),\label{eqn:perturb_rho}\\
    &P'(t,\mbf{x})=P(r)+\delta P(t,\mbf{x}),\\
    &\Phi_\ast'(t,\mbf{x})=\Phi_\ast(r)+\delta\Phi_\ast(t,\mbf{x})\label{eqn:perturb_phi},
\end{align}
where the prime symbol indicates the perturbed states and $\delta$ denotes the Eulerian perturbations. In our derivation, we will continue working with perturbations in Eulerian frames. 
For the Lagrangian approach that leads to the same result, see, \emph{e.g.}, Lynden-Bell~and~Ostriker~\cite{lyndenbell1967}.

The Poisson's equation relates $\delta \rho$ and $\delta\Phi_\ast$ in Eqs.~(\ref{eqn:perturb_rho}) and (\ref{eqn:perturb_phi})
\begin{align}
    \nabla^2 \delta\Phi_\ast =4\pi G\delta \rho.
\end{align}
In addition, from Eq.~(\ref{eqn:eos}), $\delta \rho$ and $\delta P$ are related by
\begin{align}
    &\frac{\delta \rho}{\rho}=\frac{1}{\Gamma_1}\frac{\delta P}{P} - A\bm{\chi}\cdot \hat{\mbf{r}},\\
    &\Gamma_1 \equiv \left(\frac{{\partial}\ln P}{{\partial }\ln \rho}\right)_s,\\
    &A\equiv \frac{{\rm d}\ln\rho}{{\rm d}r} - \frac{1}{\Gamma_1 }\frac{{\rm d}\ln P}{{\rm d}r},
\end{align}
where $\Gamma_1$ is the adiabatic exponent, $A$ is the Schwarzschild discriminant, and $\bm{\chi}$ is the Lagrangian displacement of the fluid element.

Replacing the equilibrium quantities with the perturbed states in Eqs.~(\ref{eqn:mass_cons}) and (\ref{eqn:momentum_cons}), we obtain
\begin{align}
&\frac{\partial\delta \rho}{\partial t} + \nabla \cdot (\rho \delta \mbf{v})=0,\label{eqn:fluid_eom1}\\
&\rho \frac{\partial \delta \mbf{v}}{\partial t}=-\nabla \delta P - \delta \rho \nabla \Phi_\ast -\rho\nabla \delta\Phi_\ast,\label{eqn:fluid_eom2}
\end{align}
where $\delta \mbf{v}\equiv {\partial }\bm{\chi}/{\partial}t$ and we have kept terms up to linear order in the perturbation. 

Eq.~(\ref{eqn:fluid_eom2}) is the equation of motion for $\bm{\chi}$, which we break down into radial and horizontal components
\begin{align}
    \bm{\chi}=\xi^{\rm r}(t,r,\theta,\phi)\hat{\mbf{r}}+\bm{\xi}^{\rm h}(t,r,\theta,\phi),
 \end{align}
where $\bm{\xi}^{\rm h}\cdot \hat{\mbf{r}}=0$. 
We seek a solution that takes the form of a harmonic oscillator, letting $\xi^{\rm r}(t,r,\theta,\phi)=\xi^{\rm r}(r,\theta,\phi)e^{-i\omega t}$. We factor out the time-dependence of $\bm{\xi}^{\rm h}$ and the perturbative quantities similarly.
Then, in terms of the redefined  time-independent quantities, Eq.~(\ref{eqn:fluid_eom2}) read
\begin{align}
&-\omega^2\rho\xi^r = - \frac{ \partial \delta P}{\partial r} - \delta \rho g - \rho \frac{\partial \delta \Phi_\ast}{\partial r},\label{eqn:radial}\\
    &-\omega^2\rho \bm{\xi}^{\rm h} = - \nabla_{\rm h} \delta P - \rho \nabla_{\rm h} \delta \Phi_\ast,\label{eqn:horizontal}
\end{align}
where $\nabla_{\rm h}$ is the horizontal gradient,
\begin{align}
    r\nabla_{\rm h}f\equiv \frac{\partial f}{\partial \theta}\hat{\bm{\theta}} + \frac{1}{\sin\theta }\frac{\partial f}{\partial\phi}\hat{\bm{\phi}}.
\end{align}

Additionally, integrating Eq.~(\ref{eqn:fluid_eom1}) with respect to $t$ and using the expression for $\bm{\chi}$ gives
\begin{align}
\rho\nabla_{\rm h}\cdot \bm{\xi}_{\rm h}=\delta\rho +\frac{1}{r^2}\frac{\partial(\rho r^2\xi^r)} {\partial r}\label{eqn:div_xih}
\end{align}
Taking divergence of Eq.~(\ref{eqn:horizontal}) and using Eq.~(\ref{eqn:div_xih}) to substitute $\nabla_{\rm h}\cdot \bm{\xi}_{\rm h}$ leads to
\begin{align}
    \omega^2\left[\delta\rho + \frac{1}{r^2}\frac{\partial (r^2\rho \xi^r)}{\partial r}\right] = -\nabla_{\rm h}^2\delta P - \rho \nabla_{\rm h}^2\delta\Phi_\ast,\label{eqn:sep_mom}
\end{align}
where $\nabla_{\rm h}^2$ is the horizontal Laplacian
\begin{align}
    r^2\nabla_{\rm h}^2f=\frac{1}{\sin\theta}\frac{\partial}{\partial\theta}\left(\sin\theta\frac{\partial f}{\partial\theta}\right) + \frac{1}{\sin^2}\frac{\partial^2 f}{\partial\phi^2}.
\end{align}
In terms of $\nabla_{\rm h}^2$, the perturbed Poisson's equation is
\begin{align} 
   \frac{1}{r^2}\frac{\partial}{\partial r}\left(r^2\frac{\partial \delta\Phi_\ast}{\partial r}\right) + \nabla_{\rm h}^2\delta\Phi_\ast =4\pi G\delta\rho. \label{eqn:poisson1}
\end{align}

As a final step, we separate the angular part from $\xi_{\rm r}$,  $\delta P$, $\delta \rho$, and $\delta\Phi_\ast $ by expanding them in terms of spherical harmonics $Y_{lm}(\theta,\phi)$, \emph{e.g.},
\begin{align}
    \xi^r(r,\theta,\phi)=\sum_{a}\xi_a^r(r)Y_{l_am_a}(\theta,\phi),
\end{align}
where $a$ runs over 3-tuples of $(\omega_a,l_a,m_a)$ or equivalently $(n_a, l_a, m_a)$. 
For each mode $a$ (omitting the subscript $a$), Eqs.~(\ref{eqn:radial}), (\ref{eqn:sep_mom}), and (\ref{eqn:poisson1}) become
\begin{align}
&\left(A+\frac{\rm d}{{\rm d}r}\right)\left(\frac{\delta P}{\rho}\right)=\frac{(\omega^2+Ag)}{r^2}(r^2\xi^{\rm r})-\frac{{\rm d}\delta \Phi_\ast}{{\rm d}r}\label{eqn:xi_1},\\
&\left(\frac{{\rm d}\ln \rho}{{\rm d}r}+\frac{\rm d}{{\rm d}r}\right)r^2\xi^{\rm r} = \frac{r^2\delta \rho}{\rho} + \frac{l(l+1)}{\omega^2}\left(\frac{\delta P}{\rho} + \delta\Phi_\ast \right),\label{eqn:xi_2}\\
    &\frac{{\rm d}}{{\rm d} r}\left(r^2\frac{{\rm d} \delta\Phi_\ast}{{\rm d} r}\right) =l(l+1)\delta\Phi_\ast+4\pi G \delta \rho r^2\label{eqn:xi_3},
\end{align}
where we have used $r^2\nabla_{\rm h}^2Y_{lm}=-l(l+1)Y_{lm}$.

Using Eqs.~(\ref{eqn:xi_1})--(\ref{eqn:xi_3}), we can solve for an eigenmode defined by its eigenfrequency $\omega_a$ and spatial eigenfunction $\bm{\xi}_a$
\begin{align}
    \bm{\xi}_a(r,\theta,\phi)=\left[\xi^{\rm r}_a(r) +  r\xi^{\rm h}_a(r)\nabla_{\rm h}\right]Y_{l_am_a}(\theta,\phi),\label{eqn:eigenmode}
\end{align}
with the horizontal component $\xi^{\rm h}$ given as 
\begin{align}
    \xi^{\rm h}_a=\frac{1}{\omega r^2}\left[\frac{\delta P_a(r)}{\rho} + \delta\Phi_{\ast,a}(r)\right].
\end{align}

We numerically solve for $\xi^{\rm r}_a$, $\delta P_a/\rho$, $\delta \Phi_{\ast,a}$ using root-finding algorithms. 
We impose boundary conditions such that the solution does not diverge at the stellar center,
\begin{align}
    &\xi^{\rm r}_a\approx l\xi^{\rm h}_a\text{ at $r\approx 0$}.
\end{align}
At $r=R$, we require that the Lagrangian perturbation on the pressure vanishes,
\begin{align}
    &\left(\delta P_a + \frac{{\rm d}P}{{\rm d}r} \xi^{\rm r}_a\right)\bigg|_{r=R}=0.
\end{align}
Furthermore, at the surface, the gravitational potential and its derivative must be continuous, which implies
\begin{align}
    \left(\frac{\partial \delta\Phi_{\ast,a}}{\partial r}+\frac{l+1}{r}\delta\Phi_{\ast,a}\right)\bigg|_{r=R} = 0.
\end{align}

To impose the boundary condition at the stellar surface, we have neglected the effect of the NS atmosphere.
For further discussion on more realistic boundary conditions, we refer readers to Unno \emph{et~al.}~\cite{unno89} or Aerts~\emph{et~al.}~\cite{aerts10}.
\begin{figure}
    \centering
    \includegraphics[width=0.48\textwidth]{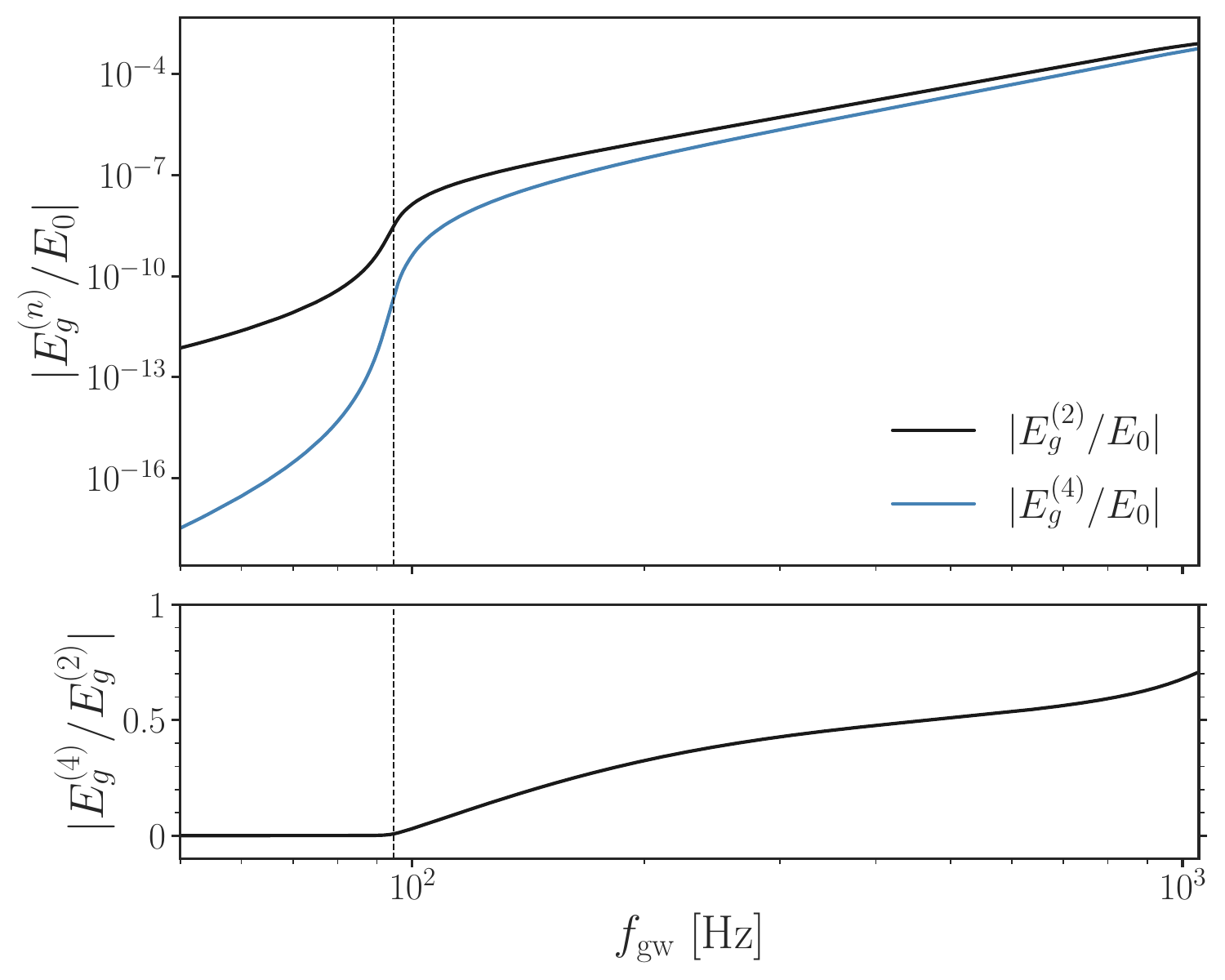}
    \caption{\textit{Top}---Comparison between the mode energies at quadratic~(black) and quartic~(blue) orders in the $n=1$, $l=|m|=2$ $g$-mode amplitudes. We normalize the energy by $E_0$. We show the linear resonance point where $2\pi f_{\rm gw}=\omega_g$~(dashed vertical). \textit{Bottom}---The ratio between the quartic and quadratic energies, which is consistently lower than unity until the merger. Thus, our analysis remains in the perturbative regime.}
    \label{fig:nonlinearity}
\end{figure}

\section{Degree of Nonlinearity\label{app:nl}}
We have modeled the RL of $g$-modes perturbatively, expanding the Hamiltonian of the star in terms of mode amplitudes.
Since RL is a nonlinear phenomenon, it is crucial to verify that our analysis is still in the perturbative regime.
Below, we compare the linear and nonlinear energies of the resonant $g$-modes. 

The linear portion that accounts for both $m=\pm 2$ $g$-modes is
\begin{align}
    &\frac{E_g^{(2)}}{E_0}=\frac{|\dot{\chi}_g|^2}{\omega_g^2} + |\chi_g|^2,
\end{align}
where the superscript (2) notes the energy is quadratic in mode amplitudes. 

As discussed in Sec.~\ref{sec:res_lock}, the quartic portion arises from (i) the formal four-mode couplings and (ii) the effective four-mode couplings chained by high-frequency modes. 
They are given by
\begin{align}
    &\Bigg[\frac{E_g^{(4)}}{E_{0}}\Bigg]_{\rm (i)}=-\frac{3}{2}\kappa_{gg^\ast gg^\ast }|\chi_g|^4,\\
    &\Bigg[\frac{E_g^{(4)}}{E_{0}}\Bigg]_{\rm (ii)}=\frac{1}{2}\sum_\beta |\chi_\beta|^2 - 4\sum_\beta^{m_\beta=0}\kappa_{gg^\ast\beta}|\chi_g|^2\chi_{\beta}^\ast\nonumber \\
    &\phantom{\Bigg[\frac{E_g^{(4)}}{E_{0}}\Bigg]_{\rm (ii)}=}-2\sum_{\beta}^{m_\beta=4}\kappa_{g^\ast g^\ast\beta}\chi_g^2\chi_\beta^\ast\nonumber \\
    &\phantom{\Bigg[\frac{E_g^{(4)}}{E_{0}}\Bigg]_{\rm (ii)}}=-2\sum_\beta^{m_\beta=0}\kappa_{gg^\ast\beta}|\chi_g|^2\chi_{\beta}^\ast-\sum_{\beta}^{m_\beta=4}\kappa_{g^\ast g^\ast\beta}\chi_g^2\chi_\beta^\ast.
\end{align}
Plugging Eqs.~(\ref{eqn:p5}) and (\ref{eqn:p7}) for the expressions for $\chi_\beta$, the energy at the quartic order in $\chi_g$ is 
\begin{align}
    \frac{E_g^{(4)}}{E_0}&=\Bigg[\frac{E_g^{(4)}}{E_{0}}\Bigg]_{\rm (i)} + \Bigg[\frac{E_g^{(4)}}{E_{0}}\Bigg]_{\rm (ii)}=-\frac{1}{2}\kappa_{\rm eff}|\chi_g|^4
\end{align}
where $\kappa_{\rm eff}$ is defined in Eq.~(\ref{eqn:k_eff_0}).

Figure~\ref{fig:nonlinearity} shows the comparison between $E_g^{(2)}$~(black) and $E_{g}^{(4)}$~(blue) as a function of $f_{\rm gw}$.
The vertical line is where $2\pi f_{\rm gw}=\omega_g$. Before RL, the energy at $E_g^{(4)}$ is negligible compared to $E_g^{(2)}$. Once RL occurs, however, the $E_g^{(4)}$ significantly increases. Nonetheless,  $|E_g^{(4)}| < |E_g^{(2)}|$ throughout the inspiral and $E_g^{(4)}$ approaches approximately $0.7E_g^{(2)}$ near the merger. Additionally, even if the inspiral lasted longer, $E_g^{(4)}$ is unlikely to exceed $E_g^{(2)}$ as the lock breaks near the merger, as seen in Fig.~\ref{fig:full_sol}. This demonstrates that our perturbative analysis is valid without the need for higher-order interaction terms.

\begin{figure}
    \centering
    \includegraphics[width=0.48\textwidth]{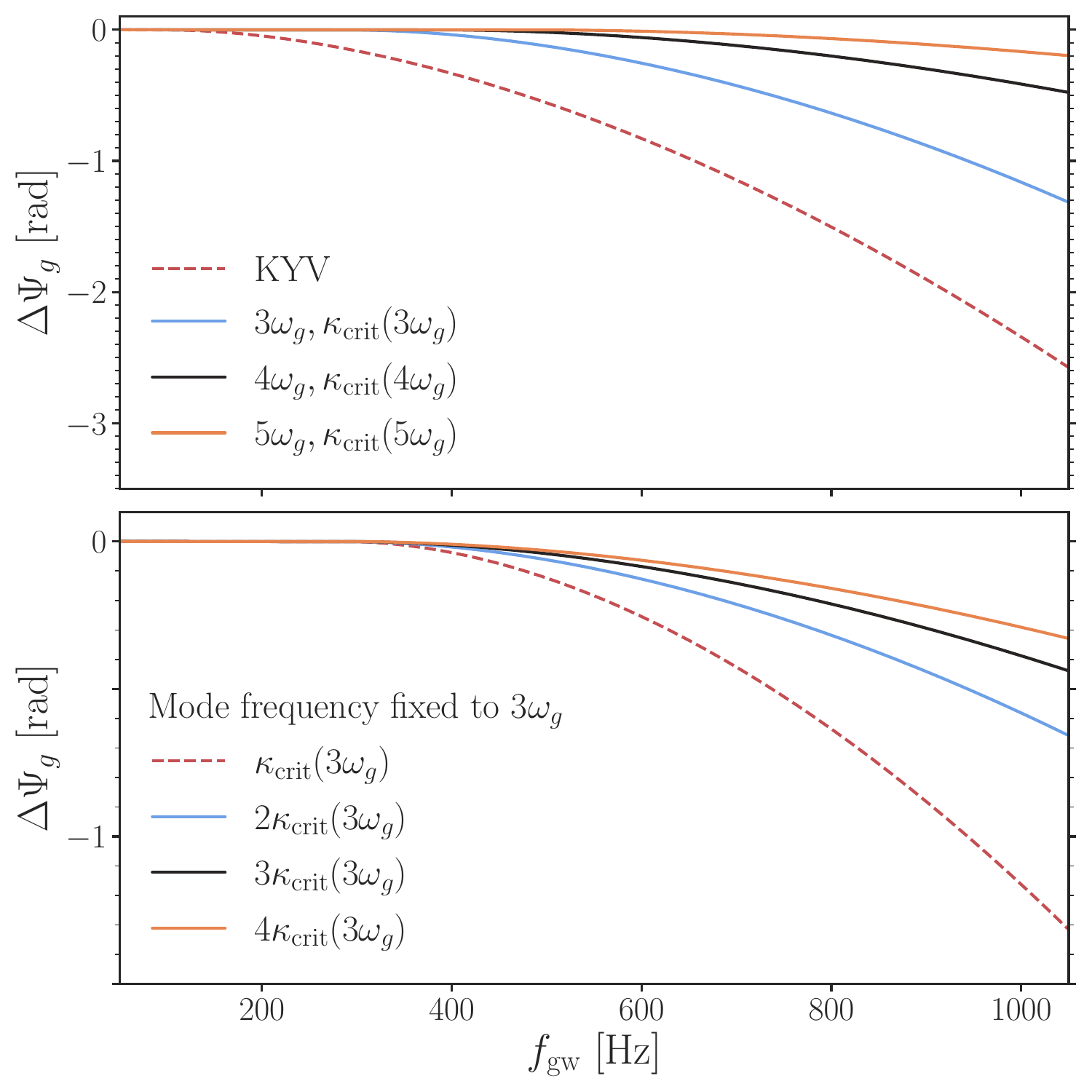}
    \caption{Impact of variations in the $g$-mode's eigenfrequency $\omega_g$ and the effective coupling coefficient $\kappa_{\rm eff}$ on the dephasing $\Delta \Psi_g$. We show $\Delta \Psi_g$ computed after multiplying $\omega_g$ by a factor of 3.5~(blue), 4~(black), and 4.5~(orange). Each curve represents the upper limit of the impact as we use $\kappa_{\rm crit}$ computed for each eigenfrequency. We show the result with original $\omega_g$ and $\kappa_{\rm eff}$~(red dashed) for comparison.}
    \label{fig:omega_var}
\end{figure}

\section{Effects of Variations in $\omega_g$ and $\kappa_{\rm eff}$}
Although our calculation of the $g$-mode's eigenfrequency $\omega_g$ and the effective coupling coefficient $\kappa_{\rm eff}$ is accurate, multiple factors such as the choices of EOS, general relativistic corrections, or superfluidity may alter their true values. 
Using Eqs.~(\ref{eqn:deph_psi})--(\ref{eqn:deph_K}), we test the impact of the variations in either $\omega_g$ and $\kappa_{\rm eff}$, assuming that the lock maintains throughout the inspiral.

In Fig.~\ref{fig:omega_var}, we demonstrate the effects of variations in $\kappa_{\rm eff}$ and $\omega_g$. 
In the top panel, we show the dephasing $\Delta \Psi_g$ computed after varying by a factor of 3~(blue), 4~(black), and 5~(orange). For the values of $\kappa_{\rm eff}$, we use $\kappa_{\rm crit}$ calculated using the adjusted frequency. 
We also show $\Delta \Psi_g$ calculated with $\omega_g$ and $\kappa_{\rm eff}$~(red dashed) for comparison. 
An increase in $\omega_g$ reduces the dephasing due to RL but still leads to $>0.1\,{\rm rad}$ at the merger in our range of variation.

In the second panel, we fix the eigenfrequency to $3\omega_g$ but vary the value of $\kappa_{\rm eff}$. 
We show the $\Delta\Psi_{g}$ for $2\kappa_{\rm crit}(3\omega_g)$~(blue), $3\kappa_{\rm crit}(3\omega_g)$~(black), and $4\kappa_{\rm crit}(3\omega_g)$~(orange). 
All our choice of $\kappa_{\rm eff}$ results in the dephasing that is $>0.1\,{\rm rad}$ near the end of the merger.

\bibliography{bib}

\end{document}